\documentclass[aps,preprintnumbers,showpacs,nofootinbib]{revtex4}
\usepackage{subfig}
\usepackage{graphicx}
\usepackage{dcolumn}
\usepackage{epsfig}
\usepackage{bm}
\usepackage{natbib}
\newcommand{\eq}{\begin{eqnarray}}
\newcommand{\en}{\end{eqnarray}}

\newcommand{\ba}[1]{\begin{eqnarray} \label{(#1)}}
\newcommand{\ea}{\end{eqnarray}}

\newcommand{\newc}{\newcommand}
\newc{\lra}{\leftrightarrow}
\newc{\beq}{\begin{equation}}
\newc{\eeq}{\end{equation}}
\newc{\barr}{\begin{eqnarray}}
\newc{\earr}{\end{eqnarray}}
  \def\vbf{\mbox{\boldmath $\upsilon$}}
\begin{document}

\topmargin -0.50in
\title { Searching for light WIMPS via their  interaction with electrons} 
\author{J.D. Vergados}
\affiliation{ University of Ioannina, Ioannina, Gr 451 10, Greece.},
\begin{abstract}
We  consider light WIMP searches  involving the detection of recoiling electrons.
\end{abstract}



\date{\today}
\begin{abstract}
In the present work we examine the possibility of detecting light dark matter particles in the few MeV region via their interactions with electrons. We analyze theoretically some key issues involved  in such a detection and  perform calculations for the expected rates, for electron recoils as well as spin induced atomic excitations, in the context of  reasonable theoretical models.
\end{abstract}
\maketitle
\section{Introduction}

The combined earlier results MAXIMA-1 \cite{MAXIMA1},\cite{MAXIMA2},\cite{MAXIMA3}, BOOMERANG \cite{BOOMERANG1},\cite{BOOMERANG2}
DASI \cite{DASI02} and COBE/DMR Cosmic Microwave Background (CMB)
observations \cite{COBE}, \cite{SPERGEL}  imply that the Universe is flat
\cite{flat01}
and that most of the matter in
the universe is Dark \cite{SPERGEL}. These results have been confirmed and improved
by the  WMAP  \cite{WMAP06} and  Planck data \cite{PlanckCP13}. Combining these data one finds:
$$\Omega_b=0.04867 \pm 0.00062, \quad \Omega _{\mbox{{\tiny CDM}}}=0.26880 \pm 0.01152 , \quad \Omega_{\Lambda}= 0.685^{+0.018}_{-0.016} $$

 On the smaller scales there exists firm indirect evidence from the
observed rotational curves, see e.g. the review \cite{UK01}, for a halo of dark matter
in galaxies and dwarf galaxies. 

Anyway in spite of the above indirect evidence for the existence of dark matter at all scales, it is essential to directly
detect such matter in order to 
unravel the nature of its constituents. 

It is clear that the direct detection of dark matter depends on the nature of the dark matter constituents and their interactions.

Historically the first dark matter particles considered  were the   WIMP's (Weakly interacting massive  particles). Given the importance of dark matter, the dominant matter component in the universe,  there is strong motivation to explore a broader set of dark matter candidates, beyond those of WIMPs, i.e. beyond candidates that interact with ordinary matter with essentially weak interactions. All such possibilities are currently examined, see  e.g    \cite{ReviewBroad17} and references there in, a   white paper summarizing the workshop “U.S. Cosmic Visions: New Ideas in Dark Matter”, which   calls out the importance of “searching for dark matter along every feasible avenue.”

Searching for WIMPs, however, still remains the most active field of research. WIMPs are  expected to have a velocity distribution with an average velocity, close to the rotational velocity $\upsilon_0=220$ km/s of the Sun around the galaxy, i.e.  they are completely non relativistic. In fact a Maxwell-Boltzmann distribution with a maximum cut off of about 2.84$\upsilon_0$  leads to a maximum energy transfer close to the average WIMP kinetic  energy  $\langle T\rangle\approx 0.4\times 10^{-6}m c^2$. Thus for GeV WIMPS this average is in the keV regime, not high enough to excite the nucleus, with only  4 exceptions \footnote{ The  exceptions are odd nuclei, with  low lying excited states, which can be populated via a Gamow-Teller like excitation, and  have  been studied theoretically \cite{VerEjSav13},  \cite{VAPSKS15} and references there in. These are: a)    transition  5/2$^+\rightarrow$7/2$^+$ to the first excited state at 57.7 keV of $^{127}$I. b)  transition  1/2$^+\rightarrow$3/2$^+$ to the first excited state at 39.6 keV  of  $^{129}$Xe. c) Transition 1/2$^+\rightarrow$3/2$^+$ to the first  excited state at 35.48 keV of  $^{125}$Te. d) Transition 9/2$^+\rightarrow$7/2$^+$ to the first  excited state at 9.4 keV of  $^{83}$Kr. Such transitions are due to the spin induced WIMP nucleus cross section. None has been observed. The same is true for the recoil experiments, which are due to the spin independent cross section}, but sufficient to measure the nuclear recoil energy.

 For light dark matter particles in the MeV region, which we will also  call WIMPs,  the average energy that can be transferred is  in the eV region.

  In the present work we will focus on light WIMPs.   WIMPs with masses below the electron mass can only be detected by special materials involving essentially free electrons, like superconductors, by measuring the total deposited energy. Heavier WIMPS with a mass less than 50 times  the electron mass can be detected  by measuring the  electron recoil, following the WIMP-electron interaction, in the case of some targets that  posses weakly bound electrons. They can also be detected by inducing atomic excitations. 

 The event rate for such a process can
be computed from the following ingredients ~\cite{LS96}: i) The elementary WIMP-electron cross section. ii)  The WIMP  density in our vicinity obtained from the rotation curves. This yields a large number density  due to the assumed smallness of the WIMP mass, expected to be about  six orders of magnitude larger than that involved in the usual WIMPs considered in nuclear recoils. iii) The WIMP velocity distribution.
In the present work we will consider a Maxwell-Boltzmann (MB) distribution in the galactic frame, with the WIMP velocity appropriately  transformed  in  the local frame.

In all  recoil experiments, like the nuclear measurements first proposed more than 30 years ago \cite{GOODWIT}, in order to overcome the formidable  background problems  one can exploit the modulation effect,  a periodic signal due to the motion of the earth around   the Sun. Unfortunately this effect, also proposed a long time ago~\cite{Druck} and subsequently studied by many authors~\cite{%
PSS88,GS93,RBERNABEI95,LS96,ABRIOLA98,HASENBALG98,JDV03,GREEN04,SFG06,FKLW11}, was found to be small
in the case of nuclear recoils. We expect it to be larger  in the case of electron recoils.

In spite of these problems many experiments undertook the task of detecting nuclear recoils in WIMP-nucleus scattering, see e.g.  \cite{XMASS09,CDMSII09,EDELWEISS11,KIMS12,SIMPLE12,PICASSO12,DAMAEPJ13,
CRESST,XENON10017,LUX14}. 
 None has been detected but very stringent limits on the nucleon cross section have been set, which can be found in a recent review\cite{KSTH18}. Furtherore projected sensitivities of Dark Matter direct detection experiments to effective WIMP-nucleus couplings have also appeared\cite{LUXZEP}.

The above results combined with theoretical motivations stimulated interest in lower mass WIMPs, see e.g. the recent work \cite{EMV12}. In fact the first direct detection limits on sub-GeV dark matter from XENON10 have recently been obtained \cite{EMMPV12}. Subsequently detection of electrons in such searches has been considered  \cite{EssVolYu17}. It is encouraging that light WIMPs in the keV  region can be detected employing Superfluid Helium  \cite{SchZur16}.

 It is, however, clear that  lighter WIMPs, with a mass of the order of that of the electron, are quite different in  energy and momentum transfer to the target. 
One, thus,  needs suitable detectors, which maybe completely different from  current WIMP detectors employed for heavy WIMP searches. 
In fact for WIMPs in the mass range of the electron mass the available energy is in the eV region and, thus, the detection of    electron recoils is possible only for electrons with very low binding energies. Therefore the detector should be able to measure recoil energy  in few eV region.

Regarding the evaluation of the elementary WIMP-electron cross section we will consider the following  possibilities:

i) Scalar WIMPs. Such particles are viable cold dark matter candidates. Their mass, as far as we know, has not been constrained by any experiment. This scalar WIMP couples with ordinary Higgs with a quartic coupling, which has been inferred  by the LHC experiments. Thus the WIMP interacts  with electrons via Higgs exchange with an amplitude proportional to the electron mass $ m_e$. In this case one gets a large kinematic enhancement of the cross section by a factor $m_e^2/m_{\chi}^2$ and, thus,   WIMPs lighter  than the electron are favored. For  WIMPs with such a small mass $m_{\chi} $, however, the energy transfer to the electron is not adequate to overcome the electron binding. So the target must consist of essentially free electrons. We will discuss the availability  of such targets below.
 
ii) For heavier WIMPs with masses up to 50 times that of the electron  we will  consider a model with a fermion WIMP interacting with the ordinary matter via  a Z-exchange.  In this case  some electrons with low binding energies can be ejected and detected by their recoils. This model, due to the axial coupling, leads to a spin interaction among  electrons. So,  once the target is immersed in a magnetic field, one can have, as we will see, a variety of $ \Delta m_s=\pm 1$  atomic spin  excitations, both within  the same shell or between spin-orbit partners, which can easily be detected.

iii) For WIMPs with masses greater than  50 times that of the electron, the electron binding is no longer a problem  and practically all electrons of the atom can be 
ejected. We will not discuss this situation in any detail.


The paper is organized as follows: In section  \ref{sec:particlmodel} we discuss the particle model employed. In section \ref{sec:freeelectrons} we study the detection of almost free electrons in special low temperature detectors, e.g. superconducting materials, which act as caloremeters. We will exploit the enhancement of the obtained rates due to the scalar nature of the WIMPs. In section \ref{sec:boundelectrons} we discuss the effect  of the electron binding on the expected rates  in the case of  experiments measuring electron recoils\footnote{We will not concern ourselves here with recently proposed  Aromatic Organic Targets \cite{BCKL19}, \cite{Collar18}  or other two-dimensional targets like those considered previously, see e.g. \cite{HKLTZ17}, \cite{DEMSY17}. The  latter type of detectors will be considered separately elsewhere \cite{KopVer19}.} in the case of WIMPs with a mass  higher than that of the electron. In section \ref{sec:atomicexcitations} we discuss the possibility  of detecting light WIMPs via atomic excitations. This can occur via the spin induced atomic transitions with excitation energy much smaller than the electron binding energy.
\section{The particle model.}
\label{sec:particlmodel}
We will consider two such models:
\subsection{Scalar WIMPs interacting with the Higgs particle via a quartic coupling.}
Scalar WIMP's can occur in particle models. Examples are i) In Kaluza-Klein theories for models involving    universal extra dimensions (for applications to direct dark matter detection  see, e.g., ~\cite{OikVerMou}). In such models  the scalar WIMPs are characterized by ordinary couplings, but they are expected to be quite massive. ii) extremely light  particles ~\cite{Fayet03}, which are not relevant to the ongoing WIMP searches iii) Scalar isodoublet particles  such  as those  considered previously in various extensions of the standard  model~\cite{Ma06} to provide some explanation for neutrino mass. Such particles   can be long lived, protected by a discrete symmetry, and it is claimed that they can be a light dark matter candidate relevant for searches in  WIMP-nucleus scattering.

\begin{figure}[!ht]
\begin{center}
\subfloat[]
{
\includegraphics[width=0.45\textwidth]{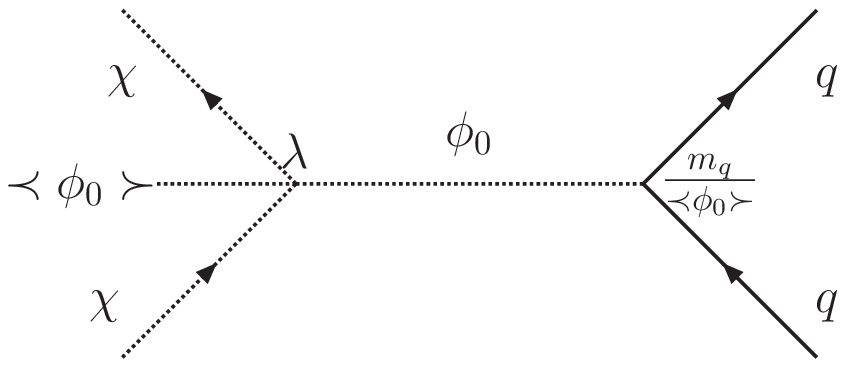}
}
\subfloat[]
{
\includegraphics[width=0.45\textwidth]{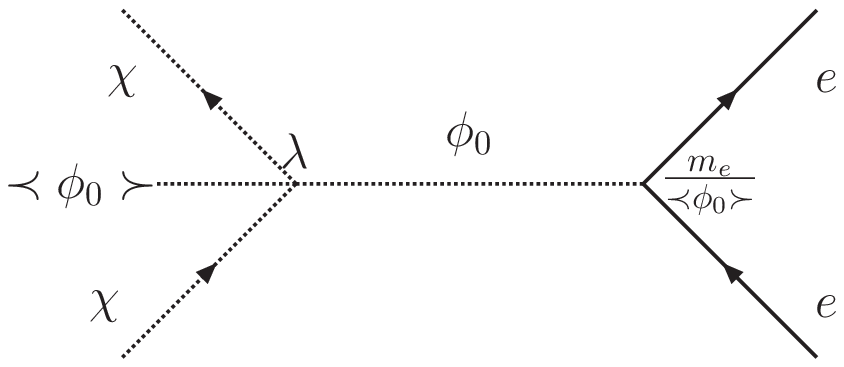}
}
\caption{(a) The quark - scalar WIMP  scattering mediated by a scalar particle, i.e. via usual Higgs exchange 
	$\phi_0 \leftrightarrow h$.  (b) The corresponding diagram for electron scalar-WIMP
scattering. Note that the relevant  amplitudes are independent of the vacuum expectation value $\langle\phi_0\rangle$ of the scalar.}
 \label{fig:xxphiphiqe}
 \end{center}
  \end{figure}

In this work we will consider  a particle model containing a  scalar particle, whose mass, to our knowledge, has not yet been constrained by any experiment. This particle, indicated by $\chi$, can be a dark matter  candidate, interacting with the neutral component $\phi_0$ of the standard model  Higgs scalar,  see Eqs (\ref{Eq:chichiphiphinew}) and (\ref{Eq:chichiphiphicross}) below, via a quartic coupling \cite{ZeeScal85,ZeeScal01,BentoRos01,BentoBero00}, and more recently  \cite{Cheung:2014pea}. It  communicates with ordinary matter via Higgs exchange, see Fig.   \ref{fig:xxphiphiqe}, and it becomes relevant for WIMP searches involving electrons.

The interest in such a WIMP has recently been revived due to a new scenario of dark matter production in bounce cosmology~\cite{Li:2014era, Cheung:2014nxi} in which the authors point out the possibility of using dark matter as a probe of a  big bounce at the early stage of cosmic evolution. 
A model independent study of dark matter production in the
contraction and expansion phases of the big bounce reveals a new venue for achieving the observed relic abundance in which dark matter was produced completely out of chemical equilibrium \cite{Cheung:2014pea}. 
In this case, this alternative route of dark matter production in bounce cosmology, can be used to test the bounce cosmos hypothesis  \cite{Cheung:2014pea}.
  
The process 
 \beq
 \phi_0+\phi_0\rightarrow \chi+\chi
 \label{Eq:chichiphiphinew}
 \eeq 
involving the scalar WIMP $\chi$ and the neutral component $\phi_0$  of the  Higgs  scalar $\phi$  proceeds  via the  quartic coupling of the Higgs potential as described by  the Feynman diagram shown in Fig. \ref{fig:xxphiphiqe}. Assuming that  the surviving component of the scalar field $\phi$ is  the Higgs $h$ discovered at the LHC, one can write down the cross section for both hadrons and electrons. The hadronic case has been studied before \cite{Cheung:2014pea} and  it  is only mentioned here for copletenes and to indicate the importance of the communication  between matter and ordinary metter via Higgs exchange. 
\\  In the case of the electron the elementary   cross section is
\beq
\sigma= \lambda^2 \frac{1}{(2 m_{\chi})^2} \frac{m_e^2}{ m^4_h}\frac{1}{2 \pi} 2 \mu^2_r=\sigma_{0H}\frac{1}{(1+x)^2}.
\label{Eq:chichiphiphicross}
\eeq
with $m_{\chi}$ and $m_h$ being the masses of the scalar WIMP and the Higgs particle respectively, $\mu_r$ the reduced mass of the WIMP-electron system, $x=\frac{m_{\chi}}{m_e}$ and
\beq 
\sigma_{0H}=\frac{1}{4 \pi}\lambda^2 \frac{m_e^2}{m_h^4}=8.4\times 10^{-45}\mbox{ cm}^2= 8.4\times 10^{-9}\mbox{pb}
\eeq


 In deriving this scale of the cross section we have assumed that the quantity $\lambda$ is the same as the quartic coupling appearing in the Higgs potential. This is determined by the LHC data, $\lambda=1/2$.  In the context of dark matter interactions it is  a rather large cross section. It is the result of the fact that, in the small Yukawa coupling $f=\frac{m_e}{\langle \phi_0\rangle}$, the vacuum expectation value $\langle \phi_0\rangle$ is canceled by that appearing in the quartic coupling. We thus emphasize that  the cross section  does not suffer from the suppression expected in the decay $h\rightarrow e^{-}e^{+}$ in which $f$ appears and, thus, it cannot be constrained by the LHC data. To the best of our knowledge it is not constrained by any other data.
\subsection{Fermion WIMPs interacting via Z-exchange.}	
Such a mechanism has been considered in the case of the lightest supersymmetric particle (LSP) for the spin induced  hadron cross section and more recently in the WIMP electron scattering \cite{VMCEKL18}. The resulting cross section depends on the coupling of the dark neutral fermions to the Z-boson, i.e. it depends on the nature of the standard model (SM) fermion and the nature of the dark matter:
\beq
{\cal L}= \frac{1}{2 \sqrt{2}}G_F J_{\mu}^Z(\chi)J^{Z\mu}(f)= \frac{1}{2 \sqrt{2}}G_FK_{\mu}(\bar{f}\gamma^{\mu}(g_V-g_A\gamma_5)f.
\eeq
In the above expression $K_{\mu}=g_{\chi}\left (\bar{\chi}\gamma_{\mu}(1-\gamma_5)\chi\right )$  stands for the axial coupling of the WIMP to the Z boson, analogous to V-A of ordinary matter. We are interested in the axial current component, since the Fermi-like coupling of the electron vanishes. We will assume further that the strength of axial current  is unity $g_{\chi}=g_V=g_A=1$. Then the invariant amplitude squared takes the form:
$${\cal M}^2=\frac{1}{8}G^2_Fg_A^2$$
Before proceeding further we will estimate the elementary WIMP-electron cross section for non relativistic electrons:
\beq
d \sigma= \frac{1}{\upsilon} \frac{1}{8}G^2_F  q^2 dq d \xi \delta(q\upsilon \xi-\frac{q^2}{2\mu_r}),\,\mu_r=\mbox{reduced mass of the WIMP electron system}, 
\eeq
(see Eq. (\ref{Eq.difsigma}), section 	\ref{sec:freeelectrons} for a kinematical derivation). Here $\upsilon$ is the velocity of the oncoming WIMP, $q$ is the momentum transfer to the  electron and $\xi=\hat{\upsilon}\cdot \hat{q}$. This  leads to the total cross section:
\beq
\sigma_e= \frac{1}{8}G^2_F \frac{1}{\pi}\mu_r^2= \frac{1}{8}G^2_F \frac{1}{\pi}m_e^2\frac{x^2}{(1+x)^2}=
\sigma_{0Z}\frac{x^2}{(1+x)^2}
\eeq
with
\beq
\sigma_{0Z}=\frac{1}{8}G^2_F \frac{1}{\pi}m_e^2\approx1.0 \times 10^{-9}
\label{Eq:sigma0z}
\eeq

It may be interesting to mention that one can infer the electron  cross section from information on  the corresponding   nucleon cross section, which  has been constrained by the  WIMP-nucleus scattering for a WIMP mass, e.g.  of 2 GeV, i.e. $\mu_r=\frac{2}{3}m_p$, by the CRESST-TUM40 experiment \cite{CRESSTTUM40}. Such a phenomenological analysis is not, however, reliable,  since the $\mu_r$ involved is  much larger. In any case, it  yields a cross section which is only a factor of three larger compared to that of Eq. \ref{Eq:sigma0z} obtained theoretically.
%
  
In this work, since  $\sigma_{0H}$ and $\sigma_{0Z}$ do not differ much,  for  simplicity and to make easier a comparison of the dependence of the obtained results on other important features 
    of the models, we will assume a common elementary cross section $\sigma_{0}$ for both Higgs and $Z$ exchange, which the average of the two.
\beq
\sigma_{0Z}\approx\sigma_{0H}=\sigma_{0}=4.0\times 10^{-45}\mbox{cm}^2=4.0\times 10^{-9}\mbox{pb}.
\eeq
In any case this  does not significantly affect the conclusions of the paper  and, if necessary,  one can re-scale the obtained rates.

	\section{The  WIMP-electron rate for free electrons}
	\label{sec:freeelectrons}
	The evaluation of the rate proceeds as in the case of the standard WIMP-nucleon scattering, but we will give the essential ingredients here to establish notation. We will begin by examining the case of a free electron.
	
 i) The case of the scalar WIMPs (SW):\\
The differential cross section, when all particles involved are non relativistic and the initial electron is at rest, can be cast in the form:
	
	\beq
d \sigma=\frac{1}{\upsilon} \lambda ^2 \frac{1}{(2 m_{\chi})}^2 \frac{m_e^2}{ m_h^4}\frac{1}{(2 \pi)^2} d^3{\bf p}'_{\chi}d^3{\bf q} \delta({\bf p}_{\chi}-{\bf p}'_{\chi}-{\bf q})\delta\left (\frac{{\bf p}^2_{\chi}}{2 m_{\chi}}-\frac{{\bf p}'^2_{\chi}}{2 m_{\chi}}-\frac{{\bf q}^2}{2m}\right ),
\eeq
where $ \upsilon$ is the velocity of the oncoming WIMP. The factor $1/(2 m_{\chi})^2$ is the usual normalization for the scalar particles and $m_h\approx 126$ GeV the mass of the exchanged Higgs particle. ${\bf p}_{\chi}$, ${\bf p}'_{\chi}$  and ${\bf q}$ are the momenta of the oncoming and outgoing WIMP and the recoiling electron respectively. The last $\delta$ function expresses the energy conservation, since the participating  particles  are non relativistic. Integrating over the momenta we find:
\beq
d \sigma=\frac{1}{2}\sigma_{0H}\frac{1}{ m_{\chi}^2} \frac{1}{\upsilon} q^2 dq d \xi \delta(q\upsilon \xi-\frac{q^2}{2\mu_r}),\,\mu_r=\frac{m_e m_{\chi}}{m_e+m_{\chi}}=\mbox{reduced mass}, 
\eeq
 
	From the energy conserving $\delta$ function one finds  that the momentum ${\bf q}$ transferred to the electron  is given by $$q=2 m_r \upsilon,\,\upsilon=\mbox{ WIMP velocity},\, \xi=\hat{\upsilon}\cdot \hat{q}\geq 0$$
	Integrating over $\xi$ with the use of the delta function one finds :
	\beq
	d\sigma=\sigma_0\frac{1}{\upsilon^2}\frac{1}{2 m_{\chi}^2 } m_e dT=\sigma_0\frac{1}{2\upsilon^2}\frac{1}{x^2}\frac{dT}{m_e},\,x=\frac{m_{\chi}}{m_e},\,\sigma_0=\sigma_{0H},
	\label{Eq.difsigma}
	\eeq
	where $T$ is the kinetic energy of the outgoing electron given by:
	\beq
	T=\frac{q^2}{2m_e}=2 \frac{1}{m_e}\mu^2_r \upsilon^2 \xi^2=2 m_e\frac{m_{\chi}^2 }{m_e^2+m_{\chi}^2 }\upsilon^2 \xi^2=2 m_e\upsilon^2 \xi^2\frac{x^2}{(1+x)^2}
	\label{Eq:Ttransf}
	\eeq
	
ii) The case of the fermion WIMP (FW).\\
Proceeding as above we find
	\beq
	d\sigma=\sigma_{0Z}\frac{1}{2\upsilon^2} \frac{dT}{m_e}
	\label{Eq.difsigma1}
	\eeq

	We are now going to discuss some parameters, which depend only on the mass of the WIMP and the velocity distribution. These are the maximum and the average electron energy. Their knowledge  provides a  qualitative understanding of the results expected from the detailed calculation.\\
	From Eq. (\ref{Eq:Ttransf}) we find  that the fraction of the energy  of the WIMP  transferred to the electron  is
	\beq
	\frac{T}{K_{\chi}}=\xi^2 \frac{x}{(1+x)^2},\,x=\frac{m_{\chi}}{m_e},
	\label{Eq:Ttransfratio}
	\eeq
	where $K_{\chi}=\frac{1}{2}m_{\chi}\upsilon^2=\frac{1}{2}m_{e}\upsilon^2x$ is the kinetic  energy of the oncoming  WIMP. We see that the maximum fraction occurs when $x=1$.

\begin{figure}[!ht]
  \begin{center}
\rotatebox{90}{\hspace{5.0cm} $T\rightarrow$ eV}
\includegraphics[width=0.7\textwidth,height=0.4\textwidth]{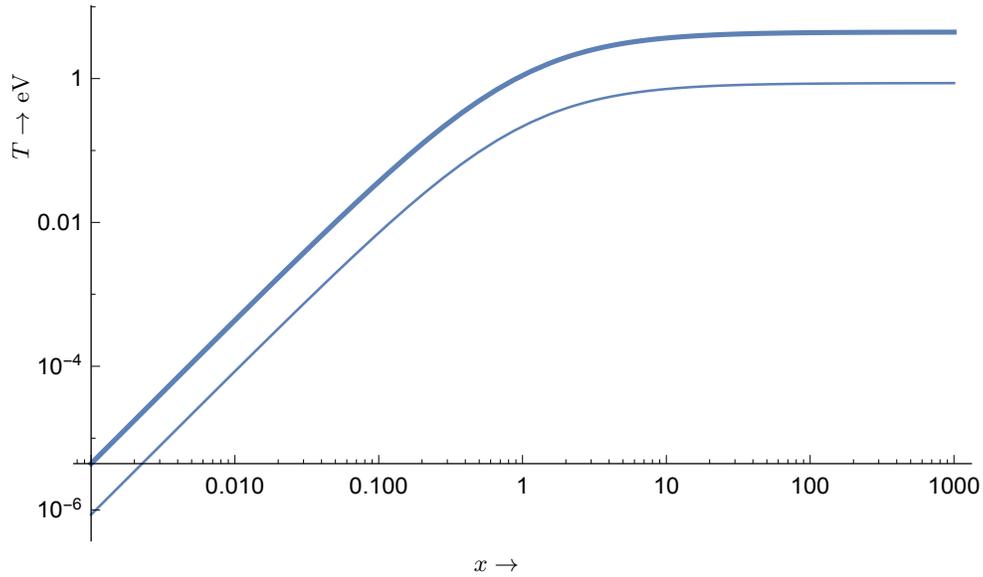}\\
\hspace{-0.0cm}$x \rightarrow $  \\
 \caption{The  maximum (thick solid line) and the average (fine solid line) energy transfer in eV  as a function of $x=\frac{m_{\chi}}{m_e}$ in the case of a free electron.}
 \label{fig:TmaxTav}
 \end{center}
  \end{figure}

 The maximum energy transfer is
 $$ T_{max}=2 m_e \upsilon^2_{esc}\frac{x^2}{(1+x)^2}$$
 i.e., in addition to $x$, it depends on the escape velocity, which is assumed to be  $\upsilon_{esc}\approx 2.84 \upsilon_0$ with  $\upsilon_0=0.7 10^{-3}c$ the Sun's velocity round the center of the galaxy. \\The electron recoiling energy depends on the direction of recoil. Its average  over all directions is $\langle T\rangle_{r}=\frac{2}{3}\upsilon^2 \frac{x^2}{(1+x)^2}$. Folding this   with the  velocity distribution, normally assumed to be of the form given by Eq.  (\ref{Eq:fyxi}) with an upper cut off equal to $\upsilon_{esc}$,   we obtain the average energy transfer $\langle T\rangle_{rv}$, which depends on $x$. The maximum and the average energy transfers $ T_{max}$ and  $\langle T\rangle_{rv}$ respectively  are exhibited in fig. \ref{fig:TmaxTav}.
 This explains why  for  WIMP mass  in the MeV region the average energy transfer is in the  range of a fraction of  eV, which is not perhaps so surprising, since, as we mentioned in the introduction, in  the earlier hadronic WIMP searches,  GeV WIMP masses  implied an energy transfer in the keV region. The    average energy can also be obtained by convoluting the  energy transfer with the differential rate (for more details see \cite{VMCEKL18}). Knowledge  the average energy is also useful in coloremetric detectors. The maximum energy affects, of course, the expected total event rate.

	 Furthermore for a given energy transfer $T$ we find:
	\beq
	\upsilon=\sqrt{\frac {m_eT}{2 \mu^2_r\xi^2}}=\left(1+\frac{1}{x}\right)\sqrt{\frac{1}{2} \frac{T}{m_e}}\rightarrow \upsilon\ge\left(1+\frac{1}{x}\right)\sqrt{\frac{1}{2} \frac{T}{m_e}}\rightarrow \upsilon_{min}=\left(1+\frac{1}{x}\right )\sqrt{\frac{1}{2} \frac{T}{m_e}}.
	\eeq
In other words the minimum velocity consistent with the energy transfer $T$ and the WIMP mass is constrained as above. The maximum velocity allowed is determined by the velocity distribution and it will be indicated by $\upsilon_{esc}$.
From this we can obtain the differential rate per electron in a given velocity volume $\upsilon^2 d \upsilon d \Omega$ as follows:
	\beq
	dR=\sigma_0\frac{\rho_{\chi}}{m_{\chi}}\frac{1}{2}\upsilon \nu(x)\frac{dT}{m_e} f({\vbf}) d\upsilon d\Omega,\,\nu(x)=\left \{ \begin{array}{cc}x^2, &\mbox{{\tiny SW}}\\ 1,&\mbox{{\tiny FW}}\\ \end{array} \right .
	\eeq
	where $ f({\vbf})$ is the velocity distribution of WIMPs in the laboratory frame. Integrating over the allowed velocity distributions we obtain:
	\beq
	dR=\frac{\rho_{\chi}}{m_{\chi}}  \sigma_0 \frac{dT}{m_e}\frac{1}{2\upsilon_0} \eta(\upsilon_{\mbox{\tiny{min}}})\times \left \{ \begin{array}{cc}{x^2}, &\mbox{{\tiny SW}}\\ 1,&\mbox{{\tiny FW}}\\ \end{array} \right .,\,\eta(\upsilon_{\mbox{\tiny{min}}})=\int_{\upsilon_{\mbox{\tiny{min}}}}^{\upsilon_{esc}} f({\vbf})\upsilon d\upsilon d\Omega
	\label{Eq:elrate1}
	\eeq
	The parameter $\eta(\upsilon_{\mbox{\tiny{min}}})$ is a crucial parameter.\\
	Before proceeding further we find it convenient to express the velocities in units of the Sun's velocity. We should also take note of the fact the velocity distribution is given with respect to the center of the galaxy. For a M-B distribution this takes the form:
	\beq
	f(y')=\frac{1}{\pi \sqrt{\pi}}e^{- y^{'2}},\,y^{'}=\frac{\upsilon^{'}}{\upsilon_0},\,\upsilon_0=220 \mbox{ km/s}
	\eeq
	We must transform it to the local coordinate system :
	\beq
	{\bf y}^{'}\rightarrow {\bf y}+{\hat\upsilon}_s+ \delta \left
(\sin{\alpha}{\hat x}-\cos{\alpha}\cos{\gamma}{\hat
y}+\cos{\alpha}\sin{\gamma} {\hat \upsilon}_s\right ) ,\,\delta=\frac{\upsilon_E}{\upsilon_0}
 \label{Eq:vlocal} \eeq 
with
$\gamma\approx \pi/6$, $ {\hat \upsilon}_s$ a unit vector in the
Sun's direction of motion, $\hat{x}$  a unit vector radially out
of the galaxy in our position and  $\hat{y}={\hat
\upsilon}_s\times \hat{x}$. The last term,  in parenthesis, in
 Eq. (\ref{Eq:vlocal}) corresponds to the motion of the Earth
around the Sun with $\upsilon_E\approx 28$ km/s being  the modulus of the
Earth's velocity around the Sun and $\alpha$ the phase of the Earth ($\alpha=0$ around June third). The above formula assumes that the
motion  of both the Sun around the Galaxy and of the Earth around
the Sun are uniformly circular. The last term in Eq. (\ref{Eq:vlocal}) containing $\delta$ is vey important in estimating the modulation effect, i.e. the time dependence of the rate.  Since $\delta$ is small we can expand the distribution in powers of $\delta$ keeping terms  up to linear in $\delta$.
\beq
			dR=\left (\frac{\rho_{\chi}}{m_{\chi}} {\upsilon_0}\right ) N_e \frac{1}{2 \upsilon_0^2}\frac{dT}{m_e}\left ( \Psi_0(y_{min})+
	 \Psi_1(y_{min}) \cos{\alpha} \right )\times \left \{ \begin{array}{cc}x^2, &\mbox{{\tiny SW}}\\ 1,&\mbox{{\tiny FW}}\\ \end{array} \right .,\,x=\frac{m_{\chi}}{m_e},
	\label{Eq:elrate2}
	\eeq
where in the above equation the first term in parenthesis represents the average  flux of WIMPs and   the second   term gives the number $N_e$ of electrons available for the scattering \footnote{In standard targets $N_e=\frac{m_t Z_{eff}}{A m_p}$, in a target of mass $m_t$ containing atoms with mass number $A$, $Z_{eff}$ represents the number of available electrons. The meaning of $Z_{eff} $ becomes clear if one takes into account that the electrons are not free but bound in the atom 
see section \ref{sec:boundelectrons}. Thus they are not all available for scattering, i.e.  $Z_{eff}<<Z $.}. Furthermore
. 
$$
y_{min}=\frac{\upsilon_{min}}{\upsilon_0}=\frac{1}{\upsilon_0}\left (1+\frac{1}{x}\right)\sqrt{\frac{1}{2} \frac{T}{m_e}},\,y_{esc}=\frac{\upsilon_{esc}}{\upsilon_0}
$$
  For a M-B distribution one finds \cite{VMCEKL18}:
\beq
\Psi_0(x)=\frac{1}{2}H\left (y_{esc}-x \right )
  \left [\mbox{erf}(1-x)+\mbox{erf}(x+1)+\mbox{erfc}(1-y_{\mbox{\tiny{esc}}})+\mbox{erfc}(y_{\mbox{\tiny{esc}}}+1)-2 \right ],\,x=y_{\mbox{\tiny min}}
  \label{Eq:Psi0MB}
\eeq
and
\barr
\Psi_1(x)&=&\frac{1}{2} H\left (y_{esc}-x \right )\delta 
   \left[\frac{ -\mbox{erf}(1-x)-\mbox{erf}(x+1)-\mbox{erfc}(1-y_{\mbox{\tiny{esc}}})-
   \mbox{erfc}(y_{\mbox{\tiny{esc}}}+1)}{2} \right . \nonumber\\
  && \left . +\frac{ e^{-(x-1)^2}}{\sqrt{\pi }}
   +\frac{
   e^{-(x+1)^2}}{\sqrt{\pi }}-\frac{ e^{-(y_{\mbox{\tiny{esc}}}-1)^2}}{\sqrt{\pi
   }}-\frac{ e^{-(y_{\mbox{\tiny{esc}}}+1)^2}}{\sqrt{\pi }}+1\right],\,x=y_{\mbox{\tiny min}}
   \label{Eq:Psi1MB}
\earr
where erf(t) and erfc(t) are the well known error function and its complement respectively.
In the above expression  the Heaviside function $H$ guarantees that the required kinematical condition is satisfied. 

After this formalism we are going to proceed in evaluating the expected spectrum of the recoiling electrons.	
The expression given by Eq. (\ref{Eq:elrate2}) can be cast in the form:
\beq
\frac{dR}{d(T/1\mbox{eV})}=\rho\Lambda \left (\Sigma_0\left (\frac{m_{\chi}}{m_e},\frac{T}{\left(1\mbox{eV}\right )}\right )+\Sigma_1\left (\frac{m_{\chi}}{m_e},\frac{T}{\left (1\mbox{eV}\right )}\right ) \cos{\alpha} \right ),\rho=\frac{1 \mbox{eV}}{2 m_e\upsilon_0^2}\approx2
\label{Eq:ediffRate}
\eeq
where 
\beq
\Sigma_i(x,s)=\frac{1}{x}\Psi_i\left (1.23\left(1+\frac{1}{x}\right )\sqrt{\rho s}\right )\times \left \{ \begin{array}{cc}{\frac{1}{x^2}}, &\mbox{{\tiny SW}}\\1,& \mbox{{\tiny FW}}\\\ \end{array} \right . ,\,i=0,1,\,s=\frac{T}{1\mbox{eV}}
\label{Eq:Sigma}
\eeq
and
\beq
	\Lambda=\frac{\rho_{\chi}}{m_e} \sigma_0 {\upsilon_0} N_e 
	\label{Eq:elrate3}
	\eeq
	Where $N_e$ the number of electrons in the target.
	
\begin{figure}[!ht]
  \begin{center}
	\subfloat[]
	{
\includegraphics[width=0.4\textwidth]{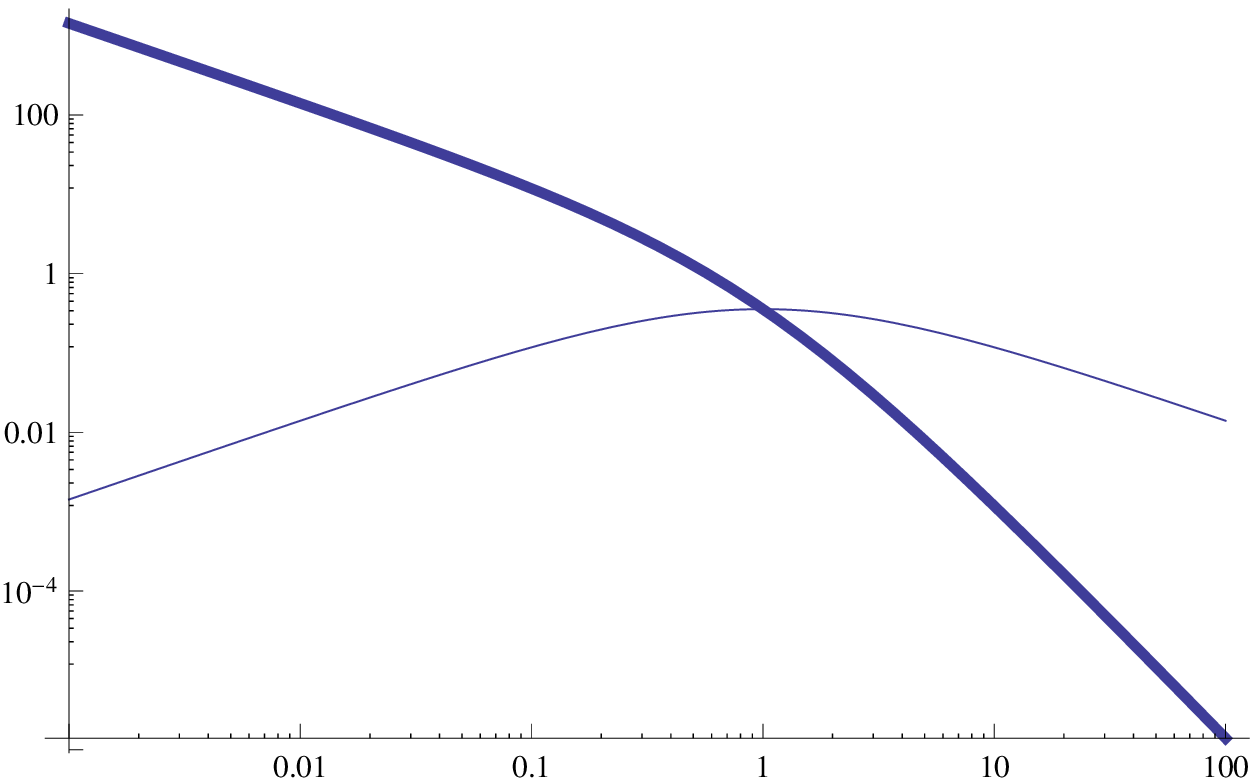}  
}
\subfloat[]
	{
\includegraphics[width=0.4\textwidth]{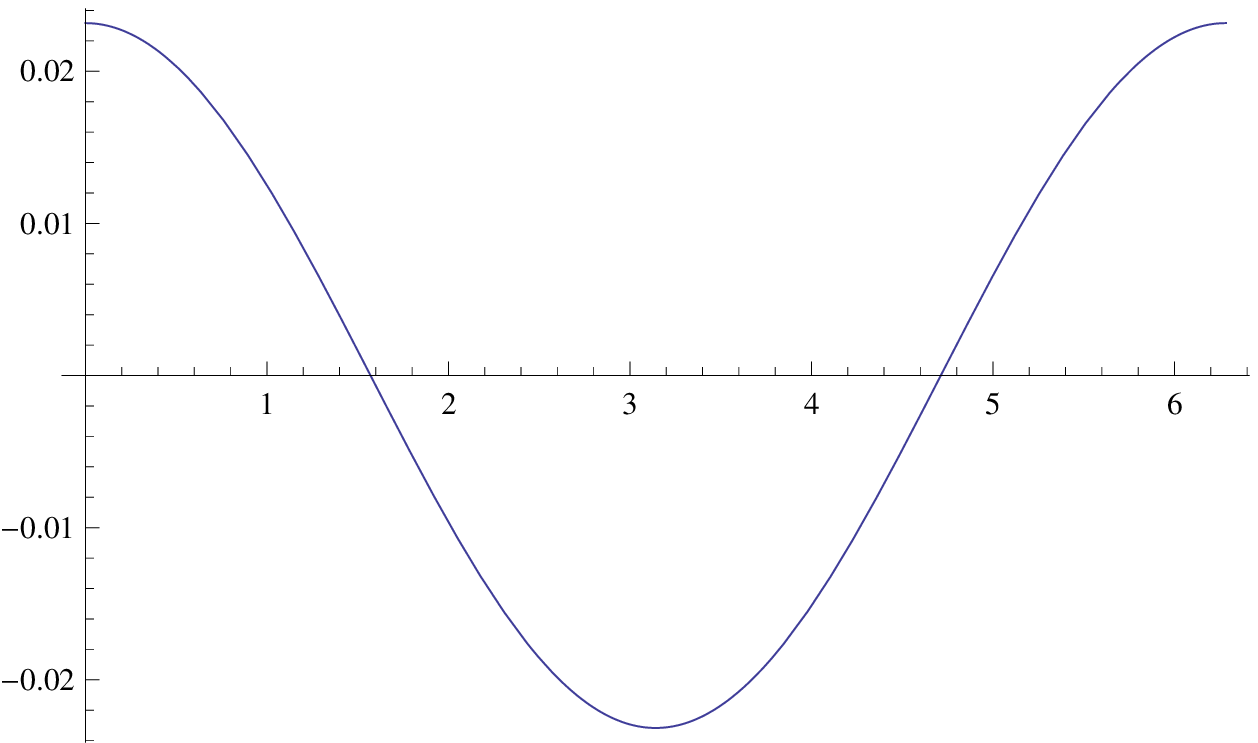}
}
\\
\subfloat[]
{
	\includegraphics[width=0.5\textwidth]{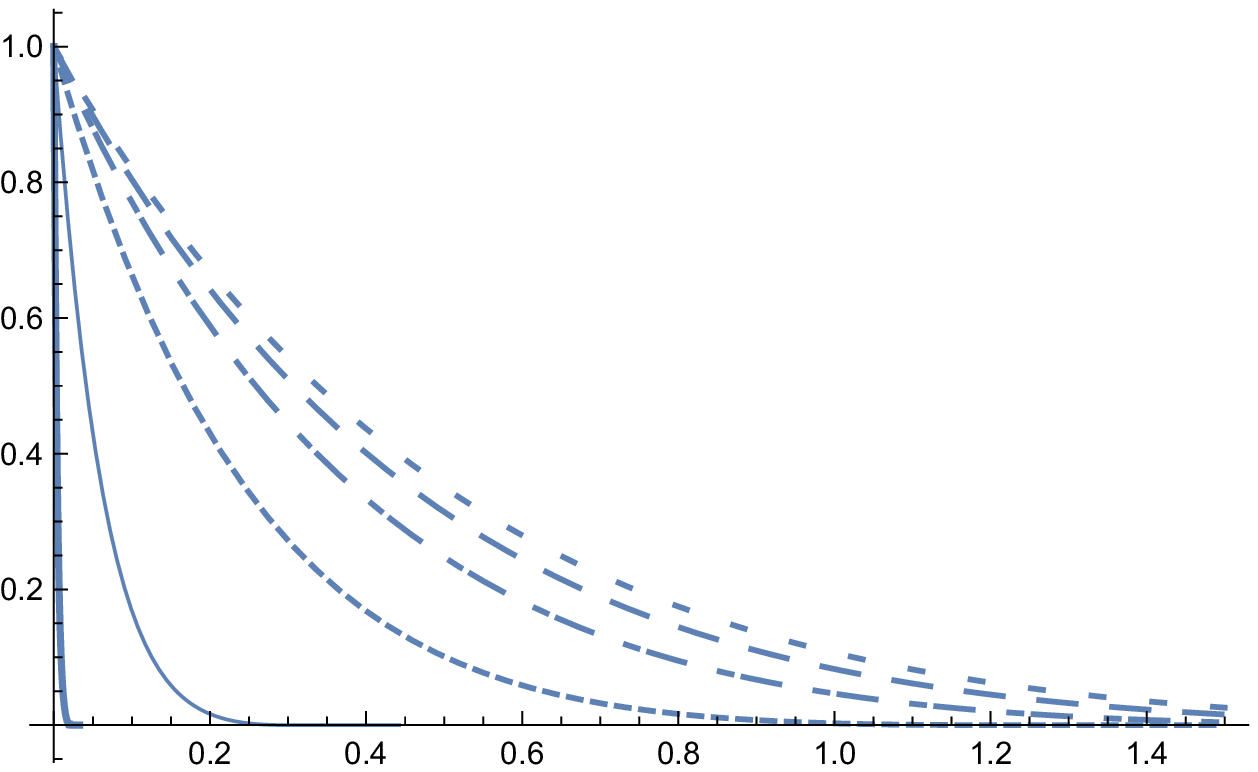} 
}\\
	\hspace{-3.0cm}$\epsilon_{th} \rightarrow eV$ 
\\
 \caption{(a) The total  time averaged event rate $R_0$,  in units of $\Lambda$ , as a function of  $x=\frac{m_{\chi}}{m_e}$. The thick   and the fine solid lines correspond to a scalar and Fermion WIMP respectively.(b)  The ratio of the time dependent to the time average rate,$\frac{R_1}{R_0}\cos{\alpha}$, as a function of the phase of the Earth $\alpha$ ($\alpha=0$ around June third).(c) The ratio of the rate with a threshold energy $\epsilon_{th}$ divided by that of zero theshold, $ R_0(\epsilon_{th})/R_0(0)$,  as a function $\epsilon_{th}$ in eV. This ratio  is the same for both scalar and Fermion WIMPs. The curves correspond to values of $x=$0.5, 1, 2, 5, 10 and 20 from left to right.}
 \label{fig:totrate}
 \end{center}
  \end{figure}

	The total  event  rates, assuming zero detector energy threshold, are  given by:
	\beq
	R_i=\Lambda \rho \int_0^{s_{\mbox{\tiny{max}}}} ds\Sigma_i(x,s),\,s_{\mbox{\tiny{max}}}=\frac{T_{\mbox{\tiny{max}}}}{\mbox{1 eV}}.
	\eeq
	The time average rate $R_0$  is exhibited in Fig. \ref{fig:totrate}a for a detector at zero anergy threshold.\\ In Fig.\ref{fig:totrate}c we show the effect of energy threshold on the rate by plotting the ratio of the rate  at a thershold energy $\epsilon_{th} $  divided by that at  zero threshold, $ R_0(\epsilon_{th})/R_0(0)$, as a function of $\epsilon_{th} $. We prefer to show this ratio rather than the individual rate  because it is independent of some parameters of the theory, e.g, the elementary  electron cross section, whether the WIMP is a scalar or Fermion etc.  \\ For the time dependence we prefer to present:
	\beq
	R_r=\frac{R_1}{R_0} \cos{\alpha},\, \alpha= \mbox{the phase of the Earth},
	\eeq
	Where $R_r$ is essentially independent of $x$ and is exhibited in Fig. \ref{fig:totrate}b.

	It is thus obvious for light WIMPs it is necessary to consider special materials in which the electrons are loosely bound, like electron pairs in a superconductor \cite{HPZ15}, provided, of course, that the number of these electrons is not very small. As another  example we mention the recently proposed  superconducting nanowires  \cite{HCNVCB19}. The latter has an energy threshold of 0.8 eV, whose  effect on the rate will be discussed below
	
	 We will, therefore, estimate the rate for free electrons, i.e.  estimate $\Lambda$ considering the following input:.

	\begin{itemize}
	\item the elementary cross section $\sigma_{0}=4\times 10^{-9}pb=4 \times 10^{-45}\mbox{cm}^2$ both for the Z and Higgs exchange.
	\item The particle density of WIMPs in our vicinity:
	$$n =0.3 \times10^3 \mbox{(MeV /cm}^3\mbox{)/0.511MeV}\approx 600\mbox{cm}^{-3}$$
	(we use the electron mass in this estimate, since  the correct mass dependence has been included through the extra factor of $1/x$ in   Eq. (\ref{Eq:Sigma})). This value leads to a flux:
	$$\Phi_0=n\times 220 \mbox{ km/s}=1.3 \times 10^{10}\mbox{cm}^{-2}\mbox{s}^{-1}=4.2 \times 10^{17}\mbox{cm}^{-2}\mbox{y}^{-1}$$
	\item The number of electrons  in the target,  estimated to be
	$$N_e=10^{24}$$
	\end{itemize}
	We thus using Eq. (\ref{Eq:elrate3}) we obtain
	$$ \Lambda\approx 1.7 \times 10^{-3}\mbox{y}^{-1}$$ 
	 From Fig. \ref{fig:totrate}a we find the time average rate at zero threshold as follows:\\ 
	\begin{itemize}
	\item $x=1\Rightarrow$\\
	 $$R_0=0.36 \times \Lambda=6. 0 \times 10^{-4}\mbox{y}^{-1}$$ 
	both for Fermion and scalar WIMPs. This is the maximum for Fermion WIMPs.
	\item For scalar WIMPs \\
	$$x= 10^{-2}\Rightarrow R_0=1.2 \times 10^2 \times \Lambda=0.2\mbox{y}^{-1}$$
	$$x= 10^{-3}\Rightarrow R_0=1.2 \times 10^3 \times \Lambda=2\mbox{y}^{-1}$$
	\end{itemize}
We should correct these values  to take into account energy threshold effects of the detector, if necessary, according to  Fig.\ref{fig:totrate}c.
	
	We should mention, however, that the WIMP detection in calorimetric experiments is still difficult, since, in spite of the large rate in the case of scalar WIMPs, the  total amount energy deposited in the detector for such a light WIMP is very small. 
	Another important issue in the case of light WIMPs is the energy threshold of the detector.  From  Fig.\ref{fig:totrate}c we see that the threshold of 0.8 eV  encountered in the proposed experiment with superconductor nanowires \cite{HCNVCB19} can  be overcome, even for small $x$, in particular for $x \ge 2$. The presence of  threshold leads, of course, to a reduction of the expected rates.
	
	Anyway it is  encouraging that it seems possible,  as it has  recently been suggested in \cite{HPZ15},  to detect even very light WIMPS, much lighter than the electron, utilizing Fermi-degenerate materials like superconductors at low temperatures. In this case the energy required is essentially the gap energy of about $1.5 kT_c$ which is in the meV region, i.e the electrons are essentially free. These authors claim that in spite of the small energy   in the range of few meV deposited to the system, the  detection of very light WIMPs becomes feasible. Furthermore it has recently been proposed \cite{CYHC19} that diamond targets can be sensitive to both  absorption processes as well as electron recoils from dark matter scattering in the WIMP mass  range of a few  MeV.

	\section{The  WIMP-electron rate for bound electrons}
	\label{sec:boundelectrons}
	In the presence of bound electrons the WIMP mass must be quite a bit larger than  the mass of the electron, $x=\frac{m_{\chi}}{m_e}>1 $. In this case it is advantageous to consider  the $Z$-exchange. Thus the
	 differential cross section for bound electrons \footnote{ Sometimes the expression is written involving $\frac{\sigma_e}{\mu^2_r}$ As we have seen in section \ref{sec:particlmodel}, however,
	 $$\sigma_e=\sigma_{0Z} \frac{x^2}{(1+x)^2}$$
	 Thus 
	 \beq
	 \frac{\sigma_e}{\mu^2_r}=\frac{\sigma_{0Z}}{m_e^2}
	 \eeq
	 The reduced mass expression is  preferred, if the WIMP-electron cross section is extracted phenomenologically.
	 }
	  takes the form:
	 \beq
	 d \sigma=\frac{\pi}{m_e^2}\frac{1}{\upsilon}\sigma_{0Z}\left |{\cal M}({\bf q})\right |^2\frac{d^3{\bf q}}{(2\pi)^3}\frac{d^3 {\bf p}'_{\chi}}{(2\pi)^3}\frac{d^3 {\bf p}_A}{(2\pi)^3} (2\pi)^3\delta \left ({\bf p}_{\chi}-{\bf p}'_{\chi}-{\bf q} -{\bf p}_A\right ) (2 \pi)\delta \left ( \frac{{\bf p}^2_{\chi}}{2 m_{\chi}}- \frac{({\bf p}')^2_{\chi}}{2 m_{\chi}}-\frac{{\bf }q^2}{2 m_e}\right )
	 \eeq
	 where again 
 ${\bf p}_{\chi}$, ${\bf p}'_{\chi}$ are  the momenta of the oncoming and outgoing WIMPs  with mass $ m_{\chi}$ and $\upsilon$ is the velocity of the oncoming WIMP. ${\bf q}$ and ${\bf p}_{A}$ are the momentum transfer to the electron and the atom respectively. The energy transfer to the atom does not appear in the energy conserving $\delta$ function, since it is negligible. Furthermore	 $${\cal M}({\bf q})=\int d{\bf r}e^{i {\bf q}.{\bf r}}\psi_{n_r,\ell,m}({\bf r})$$
	 with $\psi_{n_r,\ell,m}({\bf r})$ the bound electron wave function coordinate  space. ${\cal M}({\bf q})$ essentially represents  
	 the overlap between the  electron bound wave function and the plane wave of the outgoing electron with momentum ${\bf q}$. It can be written as  $(2 \pi)^{3/2}\Phi_{n_r,\ell,m}(a,\bf{q})$,
    with $\Phi_{n_r,\ell,m}(a,\bf{q})$ the bound electron wave function in momentum space with  $a=\frac{\alpha Z}{n_r+\ell+1}\frac{m_e c^2}{\hbar c}$. For $\ell=0$ (s-states), which are of interest in the present work, they appear in  table \ref{tab:belwf} as $\Phi_{nr}(a,q)$.
    \\
\begin{table}
\begin{center}
\caption{ The $\ell=0$ bound electron  wave functions in momentum space. $a=\frac{\alpha Z}{n_r+\ell+1}\frac{m_e c^2}{\hbar c}$, $\alpha\approx\frac{1}{137}$.
\label{tab:belwf}}
$$
\begin{array}{|c|c|}
\hline
n_r&\Phi_{nr}(a,q)\\
\hline
0&\frac{2 \sqrt{2} a^{5/2}}{\pi 
   \left(a^2+q^2\right)^2}\\
1&\frac{4 \sqrt{2} a^{5/2}
   \left(q^2-a^2\right)}{\pi 
   \left(a^2+q^2\right)^3}\\
2&\frac{2 \sqrt{2} a^{5/2} \left(3
   a^5-10 a^3 q^2+3 a
   q^4\right)}{\pi
   \left(a^2+q^2\right)^4}\\
3&\frac{8 \sqrt{2} a^{5/2}
   \left(-a^6+7 a^4 q^2-7 a^2
   q^4+q^6\right)}{\pi 
   \left(a^2+q^2\right)^5}\\
4&\frac{2 \sqrt{2} a^{5/2} \left(5
   a^4-10 a^2 q^2+q^4\right)
   \left(a^4-10 a^2 q^2+5
   q^4\right)}{\pi 
   \left(a^2+q^2\right)^6}\\
5&\frac{4 \sqrt{2} a^{5/2} \left(-3
   a^{10}+55 a^8 q^2-198 a^6
   q^4+198 a^4 q^6-55 a^2 q^8+3
   q^{10}\right)}{\pi 
   \left(a^2+q^2\right)^7}\\
	\hline
	\end{array}
$$
\end{center}
\end{table} 

\begin{figure}[!ht]
  \begin{center}
\rotatebox{90}{\hspace{-0.0cm} $\psi^2_{n_r,\ell}(u(T))\rightarrow$}
\includegraphics[width=0.7\textwidth,height=0.4\textwidth]{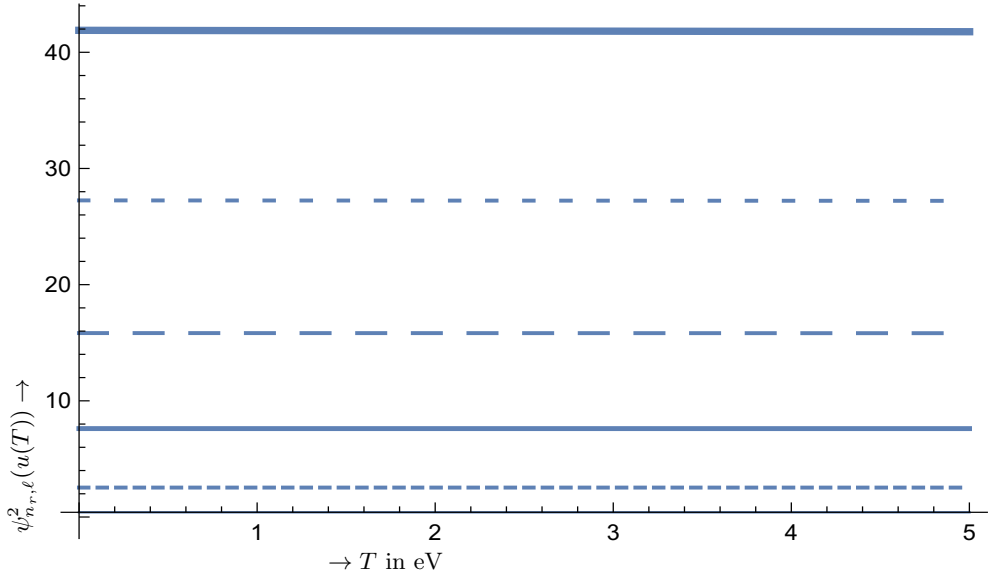}\\
\hspace{-3.0cm}$\rightarrow T$ in eV \\
 \caption{The function $\psi^2_{n_r,\ell}(u(T))$ for $\alpha Z=1/2$ is exhibited as a function of the electron recoil energy $T$ in units of eV. It is shown for $n_r=0,1,2,3,4,5$ increasing upwards (the lowest one is barely visible).}
 \label{fig:phielT}
 \end{center}
  \end{figure}

	 Thus integrating over ${\bf p}_A$ with the help of the momentum conserving $\delta$ function we obtain
	 \beq
	 d \sigma=\frac{\pi}{\upsilon}\frac{\sigma_{0Z}}{m_e^2}\frac{1}{(2 \pi)^2} \Phi_{n_r,\ell}^2(a,{\bf q})d^3 {\bf p}'_{\chi}d^3 {\bf q} \delta \left ( \frac{{\bf p}^2_{\chi}}{2 m_{\chi}}- \frac{({\bf p}')^2_{\chi}}{2 m_{\chi}}-\frac{{\bf q}^2}{2 m_e}\right )
	 \eeq 
	 Then 
	$$\int d^3 {\bf p}'_{\chi}\delta \left ( \frac{{\bf p}^2_{\chi}}{2 m_{\chi}}- \frac{({\bf p}')^2_{\chi}}{2 m_{\chi}}-\frac{{\bf }q^2}{2 m_e}\right )=4 \pi m_{\chi}^2 \upsilon\sqrt{1-\frac{2(b+T}{m_x\upsilon^2}} $$ 
	where $b$ is the binding energy of the electron and  $ T$ is the energy of the recoiling electron, $T=q^2/(2 m_e)$. Similarly  the integration over ${\bf q}$ for s-wave functions yields $ \Phi^2_{n_r,\ell}( a,\sqrt{2 m_e T}) 4 \pi \sqrt{2 m_e T} m_e dT$.
	Furthermore by writing $\sqrt{2 m_eT}=u a$ we get 
	$$\Phi^2_{n_r,\ell}( a,\sqrt{2 m_e T})=\frac{\psi^2_{n_r,\ell}(u)}{a^3}$$
	  Thus the cross section becomes
	$$
	d \sigma=\frac{4 \pi}{y}\sigma_{0Z} x^2 \psi^2_{n_r,\ell}(u) u \sqrt{y^2-\frac{2(b+T}{x m_e\upsilon_0^2}}\frac{m_e dT}{a^2},\, y=\frac{\upsilon}{\upsilon_0},
	$$
	where having in mind to eventually use the Maxwell-Boltzmann (M-B) velocity distribution we have expressed the velocity in units of $\upsilon_0=220$km/s. Measuring now the $b$ and $T$ in eV, which is the expected scale, we obtain
	\beq
	d \sigma=\frac{4\pi}{y}\sigma_{0Z} x^2\psi^2_{n_r,\ell}(u(T)) 
	\sqrt{y^2-\frac{\rho'(b+T)}{x}}  \left (\frac{n_r+\ell+1}{\alpha Z}\right )^2 \times 2 \times 10^{-6}u(T)dT,\, \rho'=3.64
	\label{Eq:dsigmau}
	\eeq
	where
	\beq
	 u(T) = \sqrt{\frac{0.2}{m_e}}\frac{n_r+\ell+1}{\alpha Z}\sqrt{T}\approx 6.3 \times 10^{-4}\frac{n_r+\ell+1}{\alpha Z} \sqrt{T}
	 \eeq
	 The behavior of the function $\psi^2_{n_r,\ell}(u(T))$ for $\alpha Z\approx \frac{1}{2}$  for various values of $n_r$ is exhibited in Fig. \ref{fig:phielT}. One can see that  the higher $n_r$ are favored. For a given $n_r$ it is essentially independent of $T$ for recoiling energies of interest to us. 
  
Returning now to Eq. (\ref{Eq:dsigmau}) we notice:

i) in folding with the velocity distribution we must integrate between $y_{min}=\sqrt{\frac{2 \rho'(b+T)}{x}}$ and $y_{esc}=2.84$

ii) for a given $x$ and $b$ the maximum electron energy is 
$$\frac{T_{max}}{\mbox{1 ev}}=\frac{y^2_{esc}x}{2 \rho' }-\frac{b}{\mbox{1 ev}}=1.1 x-\frac{b}{\mbox{1 ev}}$$
Thus for a value of $x=5$ and a binding energy 2.5 eV the maximum electron energy is expected to be 3 eV.

iii) For a given binding energy, $x$ must be at least $x_{min}=0.90 b$

Folding the cross section with the velocity distribution (see Eq. (\ref{Eq:fyxi}) below) including the extra factor of $y$ coming from the flux we obtain:
\barr
	\langle y \frac{d \sigma}{dT}\rangle&=&4\pi\sigma_{0Z} x^2 \psi^2_{n_r,\ell}\left (\frac{n_r+\ell+1}{\alpha Z}\right )^2 \times 2 \times 10^{-6}u(T)dT g(x,T,b),\nonumber\\g(x,T,b)
&=&\frac{2}{\sqrt{\pi}}\int_{y_{min}}^{y{esc}}dy y e^{-(1+y^2)}\sinh{2 y}\sqrt{y^2-\frac{\rho'(b+T)}{x}}
	\earr
The total rate can now be cast in the form
\beq
\frac{dR}{dT}=\Lambda R_{d_0},\,R_{d_0}=4\pi x \psi^2_{n_r,\ell}\left (\frac{n_r+\ell+1}{\alpha Z}\right )^2 \times 2 \times 10^{-6}u(T) g(x,T,b)
\label{Eq:dratebound}
\eeq
\beq
R=\Lambda R_0,\,R_0=4\pi x\int_0^{Tmax(x,b)} dT \psi^2_{n_r,\ell}\left (\frac{n_r+\ell+1}{\alpha Z}\right )^2 \times 2 \times 10^{-6}u(T) g(x,T,b)
\label{Eq:ratebound}
\eeq
where 
$$ \Lambda= \frac{\rho_{\chi}}{m_e}  \sigma_{0Z} \upsilon _0$$	
with  $\rho_{\chi}$ the WIMP density in our vicinity. Note that  $m_e$ rather $m_{\chi}$ has been employed in determining the number density of WIMPs with a compensating factor $1/x$  already   incorporated into Eq. (\ref{Eq:ratebound}).

There exist few atoms which possess s-state electrons with small binding energies. From atomic data tables \cite{Larkins77,Sevier72,PorFreed78} we found and  list those with $b\le 10$ eV in table \ref{tab.atomicb}. There exist, of course, states with binding energies smaller than those of the s-states, but, as we have mentioned, for light WIMPs they are not going to contribute significantly to the total rate.
\begin{table}
\begin{center}
\caption{Listed are the atoms and the  binding energy of the corresponding s-electrons. Only electrons with binding energies less than 10 eV are listed.
\label{tab.atomicb}
}  
\begin{tabular}{|c|c|c|c|c|c|c|c|}
\hline
$_{49}$In: 0.1 eV&$_{11}$Na: 0.7 eV&$_{23}$Al: 0.7 eV&$_{50}$Sn: 0.9 eV&$_{31}$Ga: 1.5 eV&$_{12}$Mg: 2.1 eV&$_{65}$Cd: 2.2 eV&
$_{82}$Pb: 3.1 eV\\$_{31}$Ge: 5.0 eV&$_{3}$Li: 5.3 eV&$_{51}$Sb: 6.7 eV&$_{14}$Si: 7.6 eV&$_{83}$Bi: 8.0 eV&$_{33}$As: 8.5 eV&$_{84}$Po: 9.0 eV&\\
\hline
\end{tabular}
\end{center}
\end{table}
It thus appears that i) NaI (b=0.7 eV in Na) as scintillator and ii)  CdTe (b=2.2 eV in Cd), Ge(Li) (b=5 eV in Ge and Li) and Si (b=7.6 eV) can be used as solid state detectors.

\begin{figure}[!ht]
  \begin{center}
\rotatebox{90}{\hspace{-0.0cm} $R\rightarrow $ events/(kg-y)}
\includegraphics[width=0.7\textwidth,height=0.4\textwidth]{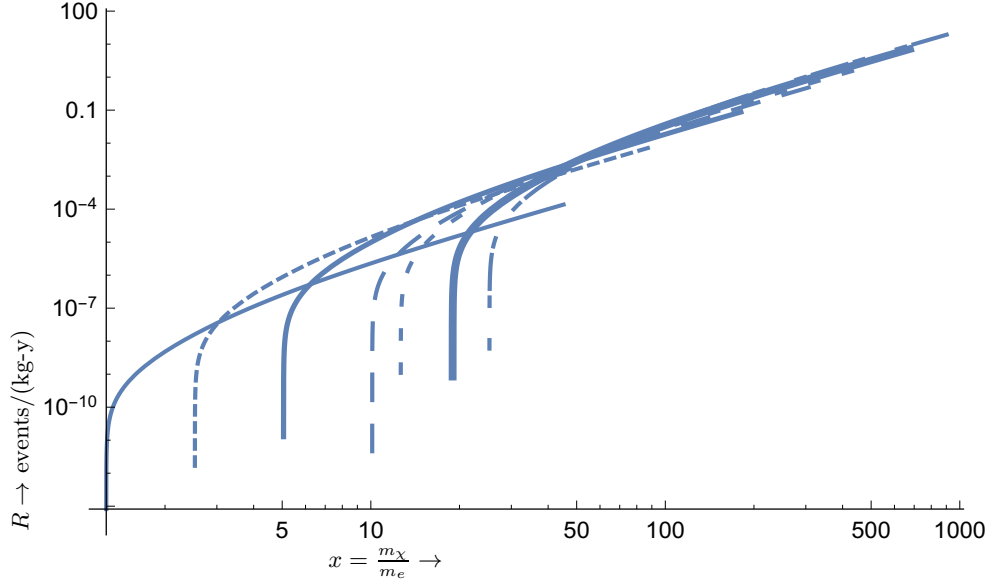}\\
\hspace{-3.0cm}$x=\frac{m_{\chi}}{m_e}\rightarrow $ \\
 \caption{The event rates as a function of $x$ for various electron binding energies $b$ in eV, which are  increasing from left to right. Thus the fine solid curve corresponds to $b=1$ $(Z_{eff}=1.0)$, the short dashed curve $b=2$ $(Z_{eff}=1)$, the intermediate thick solid line to $b=3$ $(Z_{eff}=2)$ the long  dashed curve $b=4$ $(Z_{eff}=3)$, the intermediate short dashed curve $b=6$ $(Z_{eff}=4)$, the very thick  solid line to  $b=10$ $(Z_{eff}=6)$ and the short-long dashed curve to b=15 $(Z_{eff}=8)$. One can clearly see the threshold values of $x$ for a given binding energy $b$. For illustration purposes the hydrogenic wave function with  $n_r=4,\ell=0$  has been employed. }
 \label{fig:Rate1Zef}
 \end{center}
  \end{figure}
  
  Many of the elements listed in table \ref{tab.atomicb}, involving s-electrons with low binding energies can serve as good targets, provided, of course, that recoiling electrons with energies in the few eV can be detected. Once a special target is selected, one must make an orbit by orbit calculation, based on the data of  table \ref{tab.atomicb}, and sum the cross section over all orbits multiplied with the number of electrons involved.
  
At this point we will make a simple calculation using  $N_e=N_A=10^{25}$, which corresponds to the number of atoms of a Kg of an $A=60$ target. This is an order of magnitude larger than that used in the case of free electrons  discussed in the previous section.  We thus obtain the results shown in Fig. \ref{fig:Rate1Zef}  using  $Z_{eff}$ much smaller than $Z$ for a typical atom. In spite of the larger $N_e$,  for low  $x$ the obtained results are  smaller  than those obtained in the previous section. We can trace this suppression to the electron binding energy $b$ through  atomic parameter $a$, which is of the order of $m_e$, much larger than the electron recoiling energies, which, for  $x<10$, tend to be in the few eV region.

   The results, of course, tend to further  increase  approximately linearly with $x$ and eventually, for $x>50$, electron recoils become easily detectable. For such values of $x$, of course, all electrons can participate, i.e. $Z_{eff}=Z$
	\section{Atomic excitations}
	\label{sec:atomicexcitations}
	We have seen that detecting low mass WIMPs by observing  recoiling electrons is pretty hard, since few electrons can be ejected, due to their binding in the atom. This problem does not persist, if the electrons are not ejected, but promoted to a higher level and the de-excitation photons are observed. In this case  an energy difference even much smaller than eV is available, if the target is placed  in a magnetic field at low temperature.
	
	As a matter of fact the  axial current present in the Z-mediated WIMP-electron interaction through the electron spin  can cause atomic transitions between atomic levels within states, which have the same radial quantum numbers and angular quantum numbers $j_1,m_1$ and $j_2,m_2$. If the atom is placed in a magnetic field the transition matrix element is expressed in terms of the Glebsch-Gordan coefficient and the nine- j symbol:
	\barr
	{\cal M}[(n,\ell,j_1m_1)\rightarrow(n,\ell,j_2m_2)]&=&C_{\ell,j_1,m_1,j_2,m_2},\nonumber\\
	C_{\ell,j_1,m_1,j_2,m_2}&=&\langle j_1\,m_1,1\,m_2-m_1|j_2\,m_2\rangle\sqrt{(2 j_1+1)3}\sqrt{2 \ell+1}\sqrt{6}\left \{ \begin{array}{ccc}\ell&\frac{1}{2}&j_1\\\ell&\frac{1}{2}&j_2 \\0&1&1 \end{array}\right \}
	\earr
	When $j_1=j_2$ the two states are those arising from the splitting of the degeneracy due to the Zeeman effect with an energy difference $\delta E=E_f-E_i= \mbox{ a few}\mu$eV. If $j_1\ne j_2$ the two levels correspond the spin orbit partners with energy differences in the eV region. For the readers convenience these matrix elements are tabulated for some cases of practical interest and are  given in the Appendix, see section \ref{Appendix}.
	
	The differential cross section now takes the form:
	\beq
d \sigma=\frac{1}{\upsilon} \sigma_{0Z}  \frac{\pi}{ m^2_e} \frac{1}{(2 \pi)^2}\left (C_{\ell,j_1,m_1,j_2,m_2}\right )^2 d^3{\bf p}'_{\chi}d^3{\bf q} \delta({\bf p}_{\chi}-{\bf p}'_{\chi}-{\bf q})\delta\left (\frac{{\bf p}^2_{\chi}}{2 m_{\chi}}-\frac{{\bf p}'^2_{\chi}}{2 m_{\chi}}-\delta E\right )
\eeq
where ${\bf q}$ the momentum transfer to the atom and $\delta E$ the excitation energy. The recoil energy of the atom is negligible.
 Integrating over the momentum ${\bf q}$ we find:
\beq
d \sigma=\frac{1}{\upsilon} \sigma_{0Z}\frac{\pi}{ m^2_e} \frac{1}{(2 \pi)^2 } \left (C_{\ell,j_1,m_1,j_2,m_2}\right )^2 d^3{\bf p}'_{\chi}\delta\left (\frac{{\bf p}^2_{\chi}}{2 m_{\chi}}-\frac{{\bf p}'^2_{\chi}}{2 m_{\chi}}-\delta E\right ).
\eeq
Performing the remaining integration we get
\beq
d \sigma=\frac{1}{\upsilon} \sigma_{0Z}\frac{\pi}{ m^2_e} \frac{1}{(2 \pi)^2} \left (C_{\ell,j_1,m_1,j_2,m_2}\right )^2 4 \pi \left .\frac{{{\bf p}'^2_{\chi}}}{| p'_{\chi}/m_{\chi}}\right |_{p'_{\chi}=\sqrt{p^2_{\chi}-2 m_{\chi}\delta E}}=\frac{1}{\upsilon} \sigma_{0Z}\frac{m^2_{\chi}}{ m^2_e}  \left (C_{\ell,j_1,m_1,j_2,m_2}\right )^2\sqrt{\upsilon^2-\frac{2\delta E}{m_{\chi}}}
\eeq

We must now fold it with the velocity distribution in the local frame, ignoring the motion of the Earth around the Sun, i.e.
\beq
f(\upsilon,\upsilon_0,\xi)=\frac{1}{\upsilon_0^3\pi \sqrt{\pi}}e^{-\frac{\upsilon^2+2\upsilon\upsilon_0\xi+\upsilon_0^2}{\upsilon_0^2}}
\label{Eq:fyxi}
\eeq
The integral over $\xi$ is done analytically to yield:
\beq
\langle (\sigma y)\rangle =\sigma_{0Z}  (C_{\ell,j_1,m_1,j_2,m_2})^2\frac{m_{\chi }^2}{ m^2_e}\frac{4}{\sqrt{\pi}}\int_{b}^{y_{max}} dy y y^2  e^{-y^2-1} \frac{\sinh (2y)}{2 y}
	\sqrt{1-\frac{b^2}{y^2}},\, b=\sqrt{\frac{2 \delta E}{m_{\chi}v_0^2}}
\eeq
($b$ here should not be confused with the electron binding energy) or
\barr
\langle (\sigma y)\rangle &=&\sigma_{0Z}  (C_{\ell,j_1,m_1,j_2,m_2})^2\nonumber\\& &x^2 \frac{2}{\sqrt{\pi}}\int_{b}^{y_{max}} dy  y^2  e^{-y^2-1} \sinh (2y)
	\sqrt{1-\frac{b^2}{y^2}},\, b=\sqrt{\frac{2 \delta E}{x m_{e}v_0^2}},\, x=\frac{m_{\chi}}{m_e}
	\earr
The last integral can only be done numerically.
	
	The event rate, omitting the orbit dependent angular momentum coefficient $(C_{\ell,j_1,m_1,j_2,m_2})^2$takes the form:
	\beq
	R=\frac{\Lambda}{x}   x^2\frac{2}{\sqrt{\pi}} \int_{b}^{y_{max}} dy  y^2  e^{-y^2-1} \sinh (2y)
	\sqrt{1-\frac{b^2}{y^2}},\,b=\sqrt{\frac{7.3\left(\delta E/1 \mbox{eV}\right )}{x}}
	\label{Eq:atomrate5}
	\eeq
	where $\Lambda$ is  defined as 
	\beq
	\Lambda= \frac{\rho_{\chi}}{m_{e}} \sigma_0 {\upsilon_0} N_e 
	\label{Eq:elrate3b}
	\eeq
		One can easily find that the  constraint among the parameters is 
		$$\sqrt{\frac{7.3\left(\delta E/1 \mbox{eV}\right )}{x}}<2.84\Rightarrow x>0.9\frac{ \delta E}{1 \mbox{eV}}$$
	
	The extra factor of $1/x$ in Eq. (\ref{Eq:atomrate5}) comes from the fact that the value of  $\Lambda$ employed has been evaluated with WIMP number density associated with a mass $m_e$, rather than $m_{\chi}$.
	
	\subsection{Some general trends}
\begin{figure}[!ht]
  \begin{center}
\subfloat[]
{
\includegraphics[width=0.4\textwidth,height=0.3\textwidth]{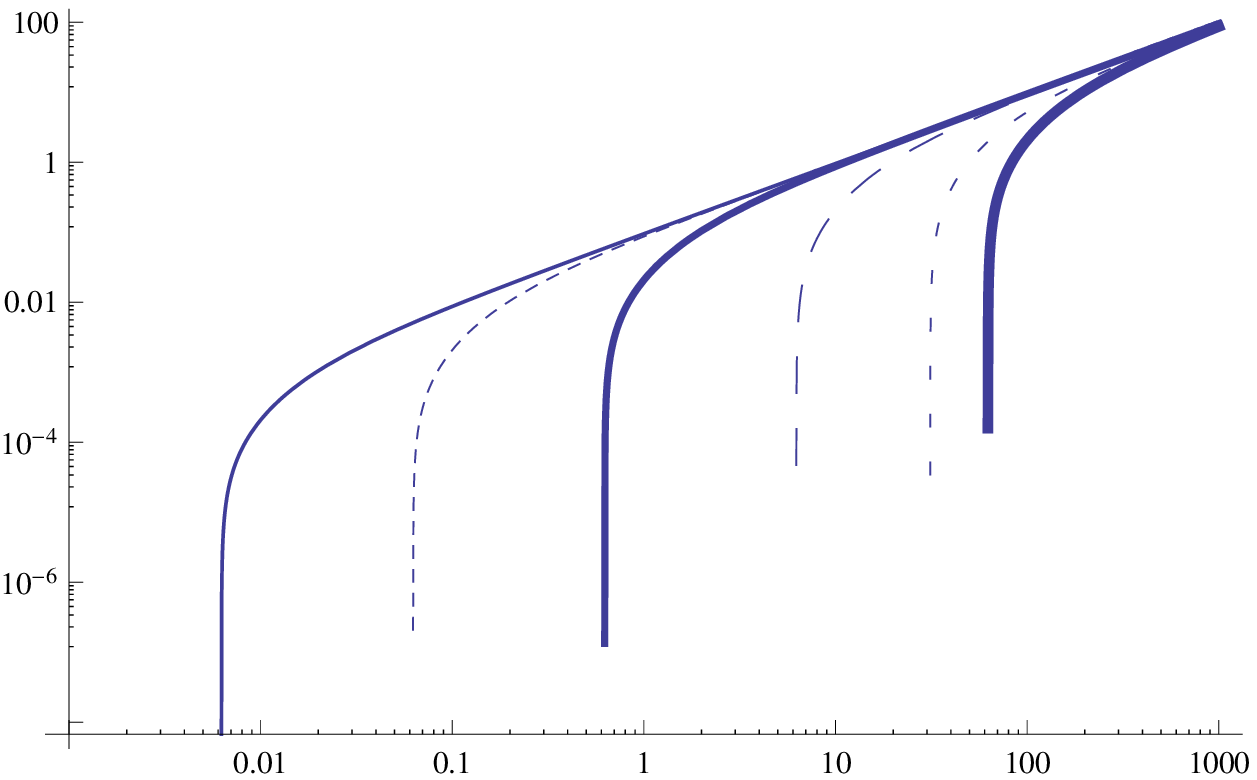}
\hspace{0.5cm}
}
\subfloat[]
{
\includegraphics[width=0.4\textwidth,height=0.3\textwidth]{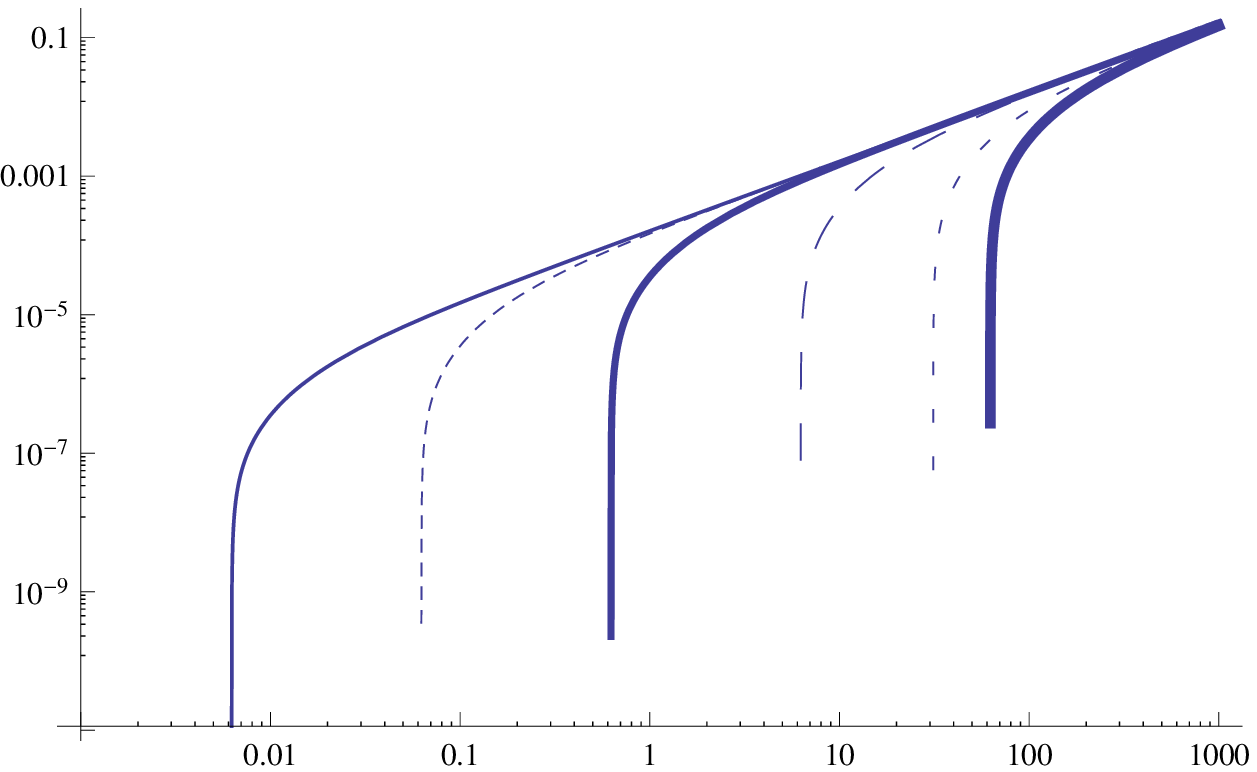}
}\\
\hspace*{-1.0cm}$\rightarrow x=\frac{m_{\chi}}{m_e}$	
 \caption{The total event rate  per year for a target with $N_A=10^{24}$ atoms as a function of  $x=\frac{m_x}{m_e}$ in the case of atomic excitations. In panel (a) the various curves correspond to   rates with $\Lambda=1$, while in panel (b) those with  the actual value of $\Lambda=1.7 \times 10^{-3}$.  In both cases the  curves correspond to values of $\delta$E= \{
   0.001,0.01,0.1,1.,5.,10.\} eV   increasing from left to right. 
 \label{fig:rateatom}}
 \end{center}
  \end{figure}

	The obtained results are exhibited in Fig. \ref{fig:rateatom}, both for $\Lambda=1$ and $\lambda=1.7 \times 10^{-3}$, assuming  one electron per atom.
	
The detection involves measuring the de-excitation of the populated level. It is also possible, following Sikivie's ideas \cite{Sikivie14} for axion detection,  to concentrate \cite{VerAvCres18} on the population of a preferred atomic level at low excitation provided that it is not otherwise occupied by electrons.  Then, assuming that it becomes occupied  due to the WIMP-electron interaction, employ a tunable laser to further excite the electrons to a preferred level. One can thus  observe the de-excitation of this preferred   level. This may require to cool system at very low temperatures and perhaps use a target,  enriched with an impurity, if necessary, so that the system  maintains an atomic structure at the necessary  low temperature.
	
	The obtained rates in Fig. \ref{fig:rateatom} are in principle detectable. It  should be noted, however, that the angular momentum factors $(C_{\ell,j_1,m_1,j_2,m_2})^2$ have not been included. They can be easily incorporated, once a target and a specific excitation pattern are selected. These can be found in  tables \ref{tab:tab1}-\ref{tab:tab2} of the Appendix, section \ref{Appendix}.

An additional advantage of the atomic experiments is the fact that targets with  a number of electrons $N_e>10^{24}$ are feasible.
\subsection{Some special targets}
We are going to examine some special examples.\\
i) First we will consider a target with the ground being a single  $p_{1/2} $   orbital, while the $p_{3/2} $ is empty. Let us suppose that the spin orbit splitting is $\epsilon_p$. In the presence of a magnetic field the m-degeneracy is removed and the ground state is  in the state $|j_1,m_1\rangle=|1/2,-1/2\rangle$. Then we have the following spin induced transitions:
$$|1/2,-1/2\rangle \rightarrow |1/2,1/2\rangle,\,|1/2,-1/2\rangle \rightarrow |3/2,-3/2\rangle, |1/2,-1/2\rangle\rangle \rightarrow |3/2,-1/2\rangle,|1/2,-1/2\rangle \rightarrow |3/2,1/2\rangle$$ 
indicated as A,B,C and D respectively.  To leading order  the spin $g_s$  factors  are $g_s=(2/3,4/3)$ for $p_{1/2} $ and $p_{3/2} $ respectively. Thus the  energies of the transitions are
$$E_x= \left\{\frac{2 \delta
}{3},\epsilon _p-\frac{5
	\delta }{3},\epsilon
_p-\frac{\delta }{3},\delta
+\epsilon _p\right\}$$
where $\delta=\mu_B  B$ with $\mu_B$ the Bohr magneton and   $B$  the magnetic field. For a  field of 1T we find $\delta=5.788\times 10^{-5}$ eV\\
A good candidate for such a transition is $_{13}$Al, involving the orbitals $2p_{1/2} $ and $2p_{3/2}$ . We find $\epsilon _p=0.65$, which in good agreement with existing tables (https://www.nist.gov/pml/atomic-spectra-database).
Thus
$$
E_x=\{0.0000386,\epsilon_p -0.0000965, \epsilon_p-0.0000193, \epsilon_p+ 0.0000579\},\,C\{2/9, 4/3, 8/9, 4/9 \},
$$
where $C$ are the corresponding spin matrix elements.

ii) Next we will consider a target with the ground being  containing a single  $d_{3/2} $   orbital, while the $d_{5/2} $ is empty. Let us suppose that the spin orbit splitting is $\epsilon_d$. In the presence of a magnetic field the m-degeneracy is removed and the ground state is  in the state $|j_1,m_1\rangle=|3/2,-3/2\rangle$. To leading order  the $g_s$ values are $g_s=(4/5,6/5)$ for $d_{3/2} $ and $d_{5/2} $ Then we have the following spin induced transitions:
$$|3/2,-3/2\rangle \rightarrow |3/2,1/2\rangle,\,|3/2,-3/2\rangle \rightarrow |5/2,-5/2\rangle, |3/2,-3/2\rangle \rightarrow |5/2,-3/2\rangle,|3/2,-3/2\rangle \rightarrow |5/2,-1/2\rangle$$ 
indicated again  as A,B,C and D respectively. Their energies are
$$E_x= \left\{\frac{8 \delta
}{5},\epsilon _d-\frac{9
	\delta }{5},\epsilon
_d-\frac{3 \delta
}{5},\epsilon _d+\frac{3
	\delta }{5}\right\}$$
Our best candidate found in the above reference is the target  $_{21}$Sc involving the  $3d3/2 \rightarrow 3d5/2$ transitions  with $\epsilon _d=0.021$ eV. 
 Other  candidates  can also be found in the same reference, e.g.:
$ \mbox{ Z=39 (Y I, 4d3/2,5/2, 0.066 eV) and Z=71 (Lu I, 5d3/2,5/2, ~0.25 eV,}$
	  where I indicates that it is a neutral atom. 
We thus find
$$ E_x=\{0.000092608,\epsilon _d-0.000104184,\epsilon _d-0.000
034728,\epsilon _d+0.000034728\},\,C=\{ 4/25, 8/5, 16/25, 4/25 \}$$
where again $C$ are the corresponding spin matrix elements.

iii) $s1/2$ states. Such states  exist in many atomic  targets. In all such cases
$$ E_x= 2\delta=1.1 \times 10^{-3},\, C=2.$$
We note the large spin matrix element.

We thus have to calculate for each target the rates
\beq
R_i=C_i R[Ex(i,x)], i=A,B,C,D,
\eeq
where R is the expression for the rate given above.

\begin{figure}[!ht]
	\begin{center}
		\subfloat[]
		{
			\includegraphics[width=0.4\textwidth]{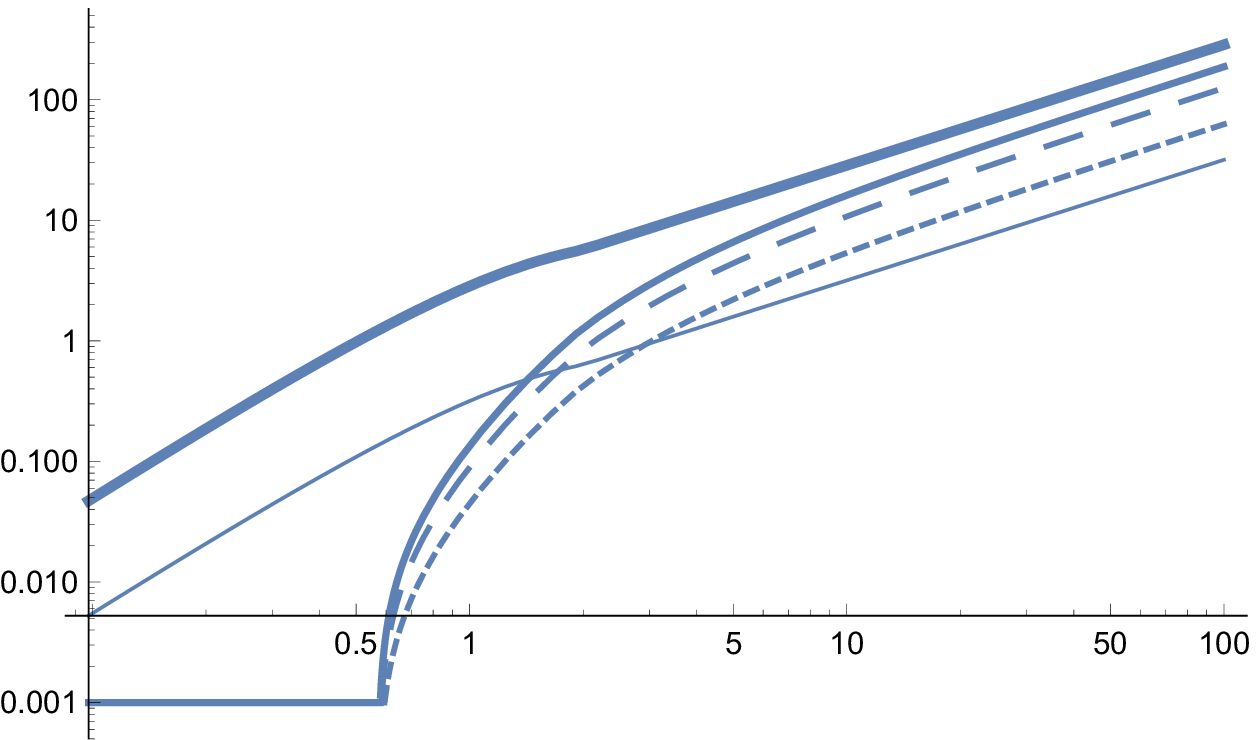}
			\hspace{0.5cm}
		}
		\subfloat[]
		{
			\includegraphics[width=0.4\textwidth]{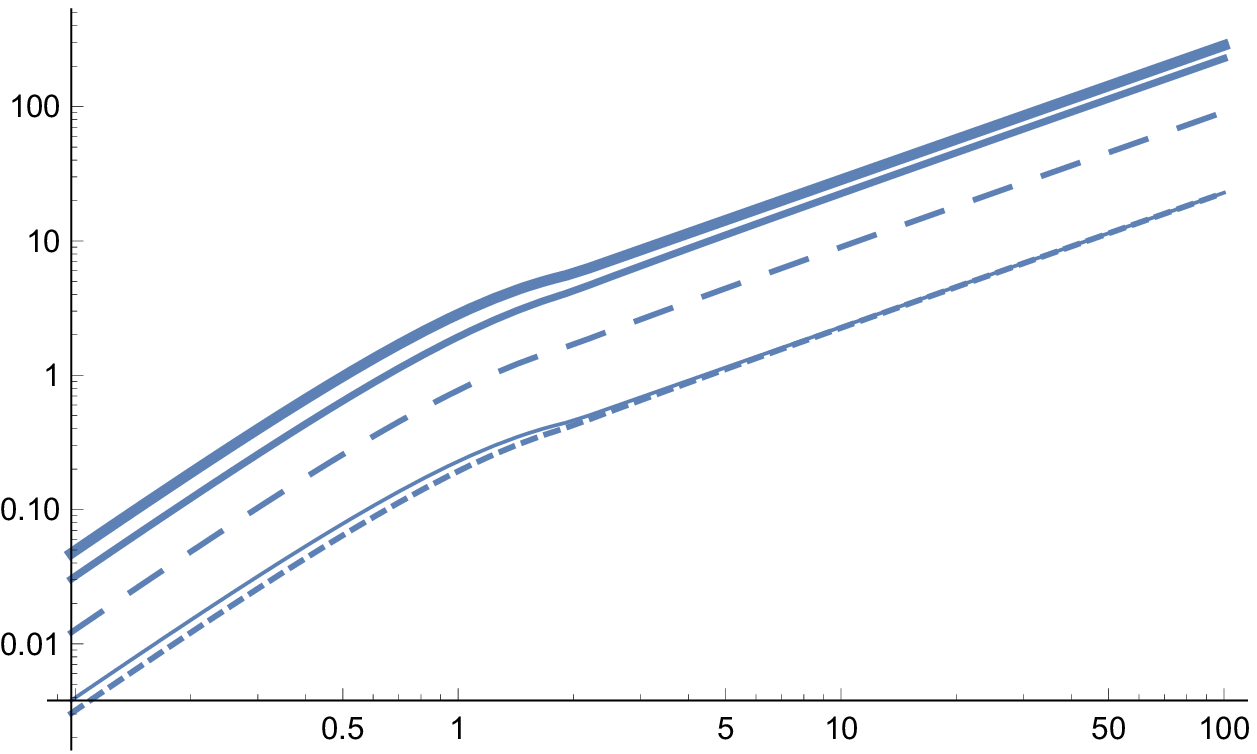}
		}\\
		\hspace*{-1.0cm}$\rightarrow x=\frac{m_{\chi}}{m_e}$	
		\caption{The total event rates, obtained  with $\Lambda=1$, for for the atomic excitations in the case of the special  targets discussed in the text. (a) In the case 2p1/2$\rightarrow$ 2p3/2 excitations in $_{13}$Al. (b) In the case of the 3d3/2$\rightarrow$ 3d5/2 excitations in $_{21}$Sc. In both cases the solid line, thick solid line, long dash and short dash correspond to the excitations of type A,B,C and D respectively (see text). For comparison we present our results for the type of s1/2 excitations, indicated by the very thick solid line. Otherwise the notation is the same with that of fig. 	\ref{fig:rateatom}. 
			\label{fig:specialatom}}
	\end{center}
\end{figure}

 The obtained results are exhibited in \ref{fig:specialatom}. Note that in the case of $s1/2$ and the A type transitions the  thresh hold value of $x$, i.e. the lowest value of the   WIMP mass required for the process to take place, is close to zero, since the spin orbit splitting does not appear.
 This also happens to be the case for
  all 3d-transitions considered here, since the spin orbit splitting is quite small (0.021 eV). On the other hand  in the case of 2p-levels  for the B,C,D transitions, a value of $m_{\chi}\ge 0.6 m_e$ is required, due to the fact that the spin orbit splitting is a bit higher (0.65 eV). 
  \\ We should not forget that the actual rates per year can be obtained after multiplying the  rates exhibited in Fig. \ref{fig:specialatom} with  $\Lambda$, $\Lambda=1.7\times 10^{-3}$ for $N=1\times 10^{24}$ atoms. 

\section{Comparison of electron recoils and atomic excitations}
In the case of light WIMPs, it is of interest to compare the electron recoil rates with those of the atomic excitation experiments. The targets are not the same, but the number of atoms in the target is taken to be the same.

 In  the case of a target with free electrons we must compare  Fig. \ref{fig:totrate}  with Figs \ref{fig:rateatom} and  \ref{fig:specialatom}.
	 The electron scattering for free electrons yields $0.36\times 1.7\times 10^{-3}=6.0 \times 10^{-4}$ events per year (y$^{-1}$),  while those associated  with 3d transitions, 
	 also in y$^{-1}$, are:
	 $$x=1 \leftrightarrow \{0.000388372,0.00388372,0.00155349,0
	 .000388372\}, $$ 
	 $$x=2 \leftrightarrow\{0.000776783,0.00776783,0.00310713,0
	 .000776783\} ,$$
	 $$x=5\leftrightarrow \{0.00194202,0.0188501,0.00753977,0.0
	 0188488\},$$ 
	 $$x=10\leftrightarrow\{0.00388407,0.0382793,0.0153115,0.00
	 382781\}, $$
	 $$ x=50\leftrightarrow\{0.0194205,0.193653,0.0774611,0.0193
	 652\} ,$$
	 $$ x=100\leftrightarrow \{0.0388411,0.387861,0.155144,0.03878
	 59\}.$$
	 The values in brackets correspond to the A, B, C, D type excitations respectively.
	 For $x=1$ the expected rates for recoil experiments are slightly higher, but for $2<x<10$ the atomic excitations  are favored.
	  
	 In the case of bound electrons we should compare the results of Figs \ref{fig:Rate1Zef} and  \ref{fig:specialatom}. We observe that in the recoil experiments,   in addition to $x$,   the rates depend on the atom, as shown in the figure, so we will only present limits on the rates for a range of $x$. For $x<5$, rates  $R\ge 10^{-6}$ y$^{-1}$ are not possible. For $5<x <10 $ the expected rates are in the range  $10^{-6}< R<10^{-5}$ y$^{-1}$. For $ 10<x<20$, $10^{-5}< R<10^{-4}$ y$^{-1}$ and for $ 20<x<30$, $10^{-4}< R<10^{-3}$y$^{-1}$.  
	 Rates of of about  $0.1$y$^{-1}$ do not appear before $x=100$.
	 \\The atomic excitation rates are  favored for  values  of $2<x<100$. 
	 
	 For  values of $x>100$ the recoil experiments are preferred, since, then, the binding electron energy becomes  unimportant and all electrons can participate..

	\section{Discussion}
	\label{sec:discussion}
	In the present paper we examined the possibility of detecting light WIMPs by exploiting their possible interactions with electrons. We calculated the event rates of various processes assuming a target with  $10^{24}$ atoms \\ For WIMPs in the mass range of the electron mass, the energy that can be transferred to the  electron is in the eV region. It is, therefore, very difficult for electrons  to escape  their binding   and be ejected. 
Detectors  utilizing Fermi-degenerate materials like superconductors  \cite{HPZ15}, have recently been suggested. In this case the energy required is essentially the gap energy of about $1.5 kT_c$ which is in the meV region, i.e the electrons are essentially free. 
The WIMP density in our vicinity becomes quite high due to their small mass and  the  WIMP-electron cross section  may be quite enhanced for scalar WIMPs. The  event rates can be reasonably high for such WIMPs, but the amount of energy deposited in the detector  is quite small. Detection of light WIMPs may become possible, even if the detectors operate with a  non zero energy threshold. Thus, e.g., in the case of  the recently proposed experiment using superconducting nanowires \cite{HCNVCB19}, the threshold of 0.8 eV can be overcome for WIMPs with $x\ge 2$

Even in the case of heavier  WIMPs, with masses up to  30 times the electron mass, only electrons with small binding can be ejected and, thus, the expected rate for electron recoils is quite small, $R\le 10^{-3}$ per year, depending on the target, see table \ref{tab.atomicb}. For still heavier WIMPs,   detection rate rises quite fast and  0.1 events per year are expected for $x=100$ and keeps rising with increasing $x$. 

We have also seen that it may be possible to detect light WIMPs via  atomic excitations due to the well known electron  spin interactions of the axial current. Thus,  using a detector at low temperature in a magnetic field, a variety of transitions between the magnetic sub-states may arise, namely $\Delta m_s= 1$ in the same shell or  $\Delta m_s=0,\pm 1$ between the spin-orbit partners. For atoms with possible  2p and 3d transitions, e.g.,  rates  up to $ 8\times 10^{-3}$ and $4\times  10^{-2}$ events per year are expected, for x=2 and $x=10$ respectively. In general  the obtained events  are higher than those expected  in the recoil experiments for $x<100$. An additional advantage is that one  can benefit  from the  very characteristic experimental signature of atomic excitations, namely the   de-excitation signals.

\section*{Acknowledgments} 
J.D.V  is happy to acknowledge support of the CERN theory division, during the last stages of this work as well as support by 
  the National Experts Council of China via  a "Foreign Master" grant while at Nanjing University, where this work began. Special thanks to Professor S. Cohen of te University of Ioannina for very fruitful discussions.

\section*{References}

\newpage
\section{Appendix}
\label{Appendix}

In this section we present the angular momentum parameters needed in evaluating the atomic excitation rates.

\begin{table}[h]
\caption{the coefficients $\left (C_{j_1,m_1,j_2,m_2,\ell}\right)^2 $ connecting via the spin operator a given initial state
 $|i\rangle=|n\ell,j_1,m_1\rangle$ with all possible states $|f \rangle=|n\ell,j_2,m_2\rangle $, for $\ell=0,\,1$. Note s-states are favored.}
\label{tab:tab1}
$$
\left(
\begin{array}{ccccc|c}
\ell&j_1&m_1&j_2&m_2&C^2_{j_1,m_1,j_2,m_2,\ell}\\
\hline
 0 & \frac{1}{2} & -\frac{1}{2} & \frac{1}{2} & \frac{1}{2} & 2 \\
\end{array}
\right),
\left (
\begin{array}{ccccc|c}
&|i\rangle&&|f\rangle&&\\
\hline
\ell&j_1&m_1&j_2&m_2&C^2_{j_1,m_1,j_2,m_2,\ell}\\
\hline
 1 & \frac{1}{2} & -\frac{1}{2} & \frac{1}{2} & \frac{1}{2} &
   \frac{2}{9} \\
 1 & \frac{1}{2} & -\frac{1}{2} & \frac{3}{2} & -\frac{3}{2} &
   \frac{4}{3} \\
 1 & \frac{1}{2} & -\frac{1}{2} & \frac{3}{2} & -\frac{1}{2} &
   \frac{8}{9} \\
 1 & \frac{1}{2} & -\frac{1}{2} & \frac{3}{2} & \frac{1}{2} &
   \frac{4}{9} \\
 1 & \frac{1}{2} & \frac{1}{2} & \frac{3}{2} & -\frac{1}{2} &
   \frac{4}{9} \\
 1 & \frac{1}{2} & \frac{1}{2} & \frac{3}{2} & \frac{1}{2} &
   \frac{8}{9} \\
 1 & \frac{1}{2} & \frac{1}{2} & \frac{3}{2} & \frac{3}{2} & \frac{4}{3}
   \\
 1 & \frac{3}{2} & -\frac{3}{2} & \frac{3}{2} & -\frac{1}{2} &
   \frac{2}{3} \\
 1 & \frac{3}{2} & -\frac{1}{2} & \frac{3}{2} & \frac{1}{2} &
   \frac{8}{9} \\
 1 & \frac{3}{2} & \frac{1}{2} & \frac{3}{2} & \frac{3}{2} & \frac{2}{3}
   \\
\end{array}
\right)
$$
\end{table}
\begin{table}[h]
\caption{The same as in table \ref{tab:tab1}, the coefficients $\left (C_{j_1,m_1,j_2,m_2,\ell}\right )^2$ for $\ell=2$}
\label{tab:tab2}
$$
\left(
\begin{array}{ccccc|c}
&|i\rangle&&|f\rangle&&\\
\hline
\ell&j_1&m_1&j_2&m_2&C^2_{j_1,m_1,j_2,m_2,\ell}\\
\hline
 2 & \frac{3}{2} & -\frac{3}{2} & \frac{3}{2} & -\frac{1}{2} &
   \frac{6}{25} \\
 2 & \frac{3}{2} & -\frac{3}{2} & \frac{5}{2} & -\frac{5}{2} &
   \frac{8}{5} \\
 2 & \frac{3}{2} & -\frac{3}{2} & \frac{5}{2} & -\frac{3}{2} &
   \frac{16}{25} \\
 2 & \frac{3}{2} & -\frac{3}{2} & \frac{5}{2} & -\frac{1}{2} &
   \frac{4}{25} \\
 2 & \frac{3}{2} & -\frac{1}{2} & \frac{3}{2} & \frac{1}{2} &
   \frac{8}{25} \\
 2 & \frac{3}{2} & -\frac{1}{2} & \frac{5}{2} & -\frac{3}{2} &
   \frac{24}{25} \\
 2 & \frac{3}{2} & -\frac{1}{2} & \frac{5}{2} & -\frac{1}{2} &
   \frac{24}{25} \\
 2 & \frac{3}{2} & -\frac{1}{2} & \frac{5}{2} & \frac{1}{2} &
   \frac{12}{25} \\
 2 & \frac{3}{2} & \frac{1}{2} & \frac{3}{2} & \frac{3}{2} &
   \frac{6}{25} \\
 2 & \frac{3}{2} & \frac{1}{2} & \frac{5}{2} & -\frac{1}{2} &
   \frac{12}{25} \\
 2 & \frac{3}{2} & \frac{1}{2} & \frac{5}{2} & \frac{1}{2} &
   \frac{24}{25} \\
 2 & \frac{3}{2} & \frac{1}{2} & \frac{5}{2} & \frac{3}{2} &
   \frac{24}{25} \\
 2 & \frac{3}{2} & \frac{3}{2} & \frac{5}{2} & \frac{1}{2} &
   \frac{4}{25} \\
 2 & \frac{3}{2} & \frac{3}{2} & \frac{5}{2} & \frac{3}{2} &
   \frac{16}{25} \\
 2 & \frac{3}{2} & \frac{3}{2} & \frac{5}{2} & \frac{5}{2} &
   \frac{8}{5} \\
 2 & \frac{5}{2} & -\frac{5}{2} & \frac{5}{2} & -\frac{3}{2} &
   \frac{2}{5} \\
 2 & \frac{5}{2} & -\frac{3}{2} & \frac{5}{2} & -\frac{1}{2} &
   \frac{16}{25} \\
 2 & \frac{5}{2} & -\frac{1}{2} & \frac{5}{2} & \frac{1}{2} &
   \frac{18}{25} \\
 2 & \frac{5}{2} & \frac{1}{2} & \frac{5}{2} & \frac{3}{2} &
   \frac{16}{25} \\
 2 & \frac{5}{2} & \frac{3}{2} & \frac{5}{2} & \frac{5}{2} & \frac{2}{5}
   \\
\end{array}
\right)
 $$
\end{table}

	\end{document}